\documentclass{emulateapj}
\usepackage{epsfig,psfig,graphics}
\usepackage{natbib}
\usepackage{apjfonts}
\usepackage{amsmath}
\usepackage{rotating}
\newcommand{\si}{$\sigma$}
\newcommand{\be}{\begin{equation}}
\newcommand{\ee}{\end{equation}}

\newcommand{\msun}{{$M_{\odot}$}}
\newcommand{\mstar}{{$M_{\star}$}}
\newcommand{\ledd}{$L_{\rm Edd}$}

\newcommand{\se}{s$^{-1}$}
 
\newcommand{\gtsima}{$\; \buildrel > \over \sim \;$}
\newcommand{\ltsima}{$\; \buildrel < \over \sim \;$}
\newcommand{\prosima}{$\; \buildrel \propto \over \sim \;$}
\newcommand{\gsim}{\lower.5ex\hbox{\gtsima}}
\newcommand{\lsim}{\lower.5ex\hbox{\ltsima}}
\newcommand{\simgt}{\lower.5ex\hbox{\gtsima}}
\newcommand{\simlt}{\lower.5ex\hbox{\ltsima}}
\newcommand{\simpr}{\lower.5ex\hbox{\prosima}}
\newcommand{\es}{erg~s$^{-1}$}
\newcommand{\esc}{erg~cm$^{-2}$~s$^{-1}$}

\newcommand{\ie}{{i.e.}}
\newcommand{\etal}{{et al.~}}
\newcommand{\cxo}{\textit{Chandra}}
\newcommand{\hst}{\textit{Hubble}}
\newcommand{\spi}{\textit{Spitzer}}
\newcommand{\bhm}{$M_{\rm BH}$}
\newcommand{\lx}{$L_{\rm X}$}

\shorttitle{AMUSE-Virgo II. Accretion Down-Sizing}
\shortauthors{Gallo et al.}

\begin{document}
\title{AMUSE-Virgo. II: Down-sizing in Black Hole Accretion}  
\author{Elena Gallo\altaffilmark{1,2}, Tommaso Treu\altaffilmark{3,4}, Philip J. Marshall\altaffilmark{5,6}, Jong-Hak Woo\altaffilmark{7}, \\Christian Leipski\altaffilmark{8}, \& Robert Antonucci\altaffilmark{3}} 
\altaffiltext{1}{MIT Kavli Institute for Astrophysics and Space Research,
70 Vassar St., Cambridge, MA 02139 } 
\altaffiltext{2}{Hubble Fellow} 
\altaffiltext{3}{Physics Department, University of California Santa Barbara, CA 93106} 
\altaffiltext{4}{Packard Fellow}
\altaffiltext{5}{Kavli Institute for Particle Astrophysics and Cosmology, Menlo Park, CA 94025}
\altaffiltext{6}{Kavli Fellow}
\altaffiltext{7}{Astronomy Program, Department of Physics and Astronomy, Seoul National University, Seoul, Korea}
\altaffiltext{8}{Max Planck Institute for Astronomy, K\"onigstuhl 17, D-69117 Heidelberg, Germany}

\begin{abstract}
We complete the census of nuclear X-ray activity in 100 early type
Virgo galaxies observed by the \cxo\ {\it X-ray Telescope} as part of
the AMUSE-Virgo survey, down to a (3$\sigma$) limiting luminosity of $3.7\times 10^{38}$ \es\ over 0.5-7 keV. The stellar mass
distribution of the targeted sample, which is mostly composed of
formally `inactive' galaxies, peaks below $10^{10}$ \msun, a regime
where the very existence of nuclear super-massive black holes (SMBHs)
is debated. Out of 100 objects, 32 show a nuclear X-ray source,
including 6 hybrid nuclei which also host a massive nuclear cluster as
visible from archival \hst\ {\it Space Telescope} images.  After
carefully accounting for contamination from nuclear low-mass X-ray
binaries based on the shape and normalization of their X-ray
luminosity function, we conclude that between $24-34\%$ of the
galaxies in our sample host a X-ray active SMBH (at the 95$\%$
C.L.). This sets a firm lower limit to the black hole occupation
fraction in nearby bulges within a cluster environment.  The
differential logarithmic X-ray luminosity function of active SMBHs
scales with the X-ray luminosity as \lx$^{-0.4\pm0.1}$ up to $10^{42}$ erg \se.  At face value, the active fraction --down to our luminosity limit-- is found to increase with host stellar mass. However, taking into account
selection effects, we find that the average Eddington-scaled X-ray luminosity scales with black hole mass as \bhm$^{-0.62^{+0.13}_{-0.12}}$, with an intrinsic scatter of $0.46^{+0.08}_{-0.06}$ dex.
This finding can be interpreted as observational evidence for `down-sizing' of black hole accretion in local
early types, that is, low mass black holes shine relatively closer to their Eddington limit than higher mass
objects. As a consequence, the fraction of active galaxies, defined as
those above a fixed X-ray Eddington ratio, {\it decreases} with increasing black hole mass.
\end{abstract}


\section{Introduction}
\label{sec:intro}

Historically speaking, active galaxies are characterized by compact
nuclei with abnormally high luminosity and fast variability ascribed to
accretion of mass onto a super-massive black hole (SMBH).  While the
term AGN (active galactic nuclei) generally refers to nuclear
luminosities in excess of $10^{43-44}$ \es, the distinction between
active and inactive is rather arbitrary, that is, is set by our
ability to detect and interpret signatures of accretion-powered
activity.  From elaboration of the Soltan argument (Soltan 1982)
follows that, since black holes have grown mostly via radiatively
efficient accretion as powerful quasars (e.g. Yu \& Tremaine 2002;
Marconi \etal 2004; Merloni \& Heinz 2008; Shankar, Weinberg \& Miralda-Escud\'e\ 2009), nearby galaxies ought to
harbor, if anything, only weakly accreting black holes.  The alleged
ubiquity of SMBHs at the center of (massive) galaxies, together with
the realization that BHs play a crucial role in regulating the
assembly history and evolution of their hosts (Kormendy
\& Richstone 1995; Kormendy \etal 1997; Magorrian et al. 1998; Gebhardt et al. 2000;
Ferrarese \& Merritt 2000; McLure \& Dunlop 2002; Marconi \& Hunt 
2003; Ferrarese \& Ford 2005; G\"ultekin \etal 2009), have spurred a
series of searches for active nuclei in the nearby universe at
different wavelengths, each with its own advantages and
limitations. In particular, the low mass end of the black hole mass
function in the local universe (Greene \& Ho 2007, 2009) remains poorly
constrained, and can only be explored indirectly, as even the highest
angular resolution attainable with current instrumentation is not
sufficient to go after a $\simlt 10^6$ \msun\ black hole through
resolved stellar kinematics, except for exceptionally nearby systems.
(see Bentz \etal 2009 on nearby, reverberation-mapped AGN).

Amongst the general population of galaxies, optical
studies suggest that nuclear activity is quite
common (43$\%$ of all galaxies in the Palomar sample; Ho, Filippenko \& Sargent 1997; Ho 2008). The percentage raises substantially in galaxies with a prominent bulge component, approaching 70$\%$ for Hubble types E-Sb. The dependence of the nuclear properties on Hubble type -- with late type objects displaying active fractions as low as $10\%$ -- has been confirmed by numerous other studies (e.g. Kauffmann et al. 2003; Miller et al. 2003; Decarli et al. 2007), although a recent work based on high-resolution mid-infrared spectrometry of a sample of (32) inactive galaxies, suggests that the AGN detection rate in late-type galaxies is possibly 4 times larger than what optical observations alone indicate (Satyapal \etal 2008).  
Since the enormous wealth of data from the Sloan Digital Sky Survey (SDSS) became
available, various environmental effects on nuclear activity have been investigated,
such as host galaxy properties (Kauffmann \etal 2003, 2004; Rich et al. 2005; 
Kewley \etal 2006; Schawinski et al. 2007, 2009; Kauffmann, Heckman \& Best 2008) local density and large scale environment (Kauffmann
et al. 2003, 2004; Constantin \& Vogeley 2006; Constantin et
al. 2008; Choi \etal 2009). 
While AGN were first discovered as powerful, unresolved optical sources at the center of galaxies, emission at higher frequencies, hard X-rays and gamma-rays, is almost univocally associated with non thermal processes related to accretion, such as Comptonization of thermal photons in a hot electron-positron plasma.  Hard X-rays in particular offer a clean-cut diagnostics, and a relatively unexplored one, to pinpoint low accretion power SMBHs in nearby galaxies. 
So far, searches for nuclear X-ray sources in formally inactive galaxies have been somewhat
sparse and focused on the high-mass end of the local population. Prior to the launch of \cxo\ and {\it XMM-Newton}, such observations were effectively limited
to X-ray luminosities $\simgt 10^{40}$ erg \se\ even
in the nearest elliptical galaxies (e.g.: Canizares, Fabbiano \& Trinchieri 1987; Fabbiano \etal 1993; Fabbiano \& Juda 1997; Allen, Di Matteo \& Fabian 2000).  Due to the lack of sensitivity and angular resolution, earlier mission were necessarily deemed to confusion between accretion-powered sources of various nature, most notably nuclear vs. off-nuclear, and also thermally emitting gas. 
As an example, 
Roberts \& Warwick
(2000), report on the detection of 54 X-ray cores out of 83 Palomar galaxies targeted by {\it ROSAT}.
As also noted by Ho (2008), a
significant (dominant, we argue) fraction of the cores' X-ray flux may be due to unresolved emission from X-ray binaries. 
The greatly improved sensitivity and angular resolution of \cxo~and {\it XMM-Newton} have made it possible to
investigate nuclear emission associated with SMBHs 
orders of magnitude deeper, effectively bridging the gap between AGN
and inactive galaxies (e.g. Di Matteo et al. 2000, 2001, 2003; Ho et al. 2001; Loewenstein et al. 2001; Sarazin, Irwin \& Bregman 2001; Fabbiano \etal 2003, 2004; Terashima \& Wilson 2003; Pellegrini 2005; Soria et al. 2006a, 2006b; Santra et al. 2007; Pellegrini, Ciotti \& Ostriker 2007; Ghosh \etal 2008; Zhang \etal 2009).

Direct measurements of bolometric Eddington ratios in {\it bona fide} AGN are typically no lower than $10^{-3}$ (e.g. Woo \& Urry 2002; Kollmeier \etal 2006; Heckman \etal 2004).
As a comparison, the inferred Eddington-scaled X-ray luminosities of inactive galaxies -- that is, of their nuclear SMBHs -- are as low as $10^{-8}$: in those massive elliptical galaxies where the temperature and density profiles of the thermally-emitting gas can be reconstructed and used to estimate the inner gas reservoir available to accretion, the measured nuclear X-ray luminosities are orders of magnitude lower than expected from Bondi-type accretion onto the nuclear SMBH (e.g. Pellegrini 2005; Soria \etal 2006a,b). 
However, most X-ray studies target massive nearby elliptical galaxies, and are thus biased towards the high mass end of the SMBH mass function. In order to expand our knowledge about black hole demographics in the local universe, it is necessary to explore both the low mass {and} the low-luminosity end of the distribution. 

Even in the nearby universe, pushing the threshold down to X-ray luminosities as low as a few $10^{38}$ erg \se\ necessarily means facing contamination from bright X-ray binaries within the instrument point spread function (PSF). 
This problem has been touched upon in a recent work by Zhang \etal (2009), who collected archival \cxo\ observations of 187 galaxies (both late and early types) within 15 Mpc. 86 of them host nuclear X-ray cores, the majority of which, based on the fitted slope of their differential luminosity function, are attributed to low-level accretion onto SMBHs, rather than to X-ray binaries. 
The issue of X-ray binary contamination becomes particularly delicate when the inferred X-ray active fractions are then used to place constraints on the local black hole occupation fraction. 
From an observational standpoint, the very existence of SMBHs in nearby dwarf galaxies remains a matter of investigation. 
Ferrarese \etal (2006a) argue that the creation of a `central massive object', be it a 
black hole or a compact stellar nucleus, would be the natural byproduct of
galaxy evolution, with the former being more common in massive bright
galaxies (with absolute B magnitude M$_{\rm B}$ brighter than $-$20), and the latter dominating
--possibly taking over-- at magnitudes fainter than $-18$ (see also Wehner \& Harris 2006, and Kormendy \etal 2009). 
Massive nuclear star clusters (e.g. Seth \etal 2008, 2010; Graham \& Spitler 2009), with inferred radii around a few tens
of pc, become increasingly prominent down the mass function. When dealing with faint X-ray cores ($10^{38}-10^{39}$ erg \se), the problem of X-ray binary contamination is further exacerbated by the presence of a nuclear star cluster, having higher stellar encounter rates, and hence a higher X-ray binary fraction with respect to the field.  
In order to deliver an unbiased census of nuclear activity for nearby galaxies down to X-ray luminosities as low as the Eddington limit for a solar mass object, not only it is mandatory to deal with nearby sources, but it also becomes necessary to have information about their stellar content within the X-ray instrument PSF, specifically the presence/absence of a nuclear star cluster. Additionally, in order to avoid contamination from the short lived, X-ray bright, high-mass X-ray binaries, deep X-ray searches for weakly active SMBHs (down to $\sim 10^{38}$ erg \se) should be preferentially limited to the nuclei of early type galaxies. \\

To this purpose, and with these caveats in mind, in Cycle 8 we proposed and were awarded a large \cxo /\spi\ program to observe 100 (84 new+16 
archival) spheroidal galaxies in the Virgo Cluster (AMUSE-Virgo: PI: Treu, 454 ks). 
The targeted sample is that of the \hst\ Virgo Cluster Survey (VCS; C\^ot\'e \etal 2004). For each galaxy, the high resolution $g$ and $z$ band images enable us to resolve, when present, the nuclear star cluster, infer its enclosed mass (following Ferrarese \etal 2006a; see Ferrarese \etal 2006b, for a detailed isophotal analysis of the \hst\ data), and thus estimate the chance contamination from a low-mass X-ray binary (LMXB) as bright/brighter than the detected X-ray core based on the shape and normalization of the LMXB luminosity function in external galaxies (see \S\ref{sec:lmxb}, and references therein). 

As a part of the AMUSE-Virgo survey, each VCS galaxy was observed with \cxo\ for a minimum of 5.4 ks, which, at the average distance of Virgo (16.5 Mpc; Mei \etal 2007; see also Blakeslee et al. 2009), yields a (3$\sigma$) sensitivity threshold of $3.75\times 10^{38}$ erg \se\ over the \cxo\ bandpass. 
The \cxo\ results from the first 32 targets (16 new + 16, typical more massive, archival observations) have been presented by Gallo \etal (2008; hereafter Paper I.). Point-like
X-ray emission from a position coincident with the optical nucleus was detected in 50 per cent of the galaxies.  We argued that, for this sub-sample, all of the 
detected nuclear X-ray sources are most likely powered by low-level
accretion on to a SMBH, with a $\simlt$11 per cent chance
contamination from LMXBs in one case only (VCC1178=NGC 4486B, for which independent evidence points towards the presence of a nuclear BH; Lauer \etal 1996). The incidence of nuclear X-ray
activity increases with the stellar mass \mstar\ of the host galaxy:
only between 3--44$\%$ of the galaxies with \mstar$<10^{10}$
\msun\ harbor an X-ray active SMBH. The fraction rises to
between 49--87$\%$ in galaxies with stellar mass above $10^{10}$
\msun. (at the 95$\%$ C.L.).

In Paper II. we complete the X-ray analysis of the whole AMUSE-Virgo sample: 
the final deliverable product of this study is an unbiased census of accretion-powered luminosity in a galaxy cluster environment, and thus the
first measurement of the SMBH activity duty cycle. 
The Paper is organized as follows: \S~\ref{sec:data} summarizes our analysis of the new \cxo\ data and the stacking procedure. In
\S~\ref{sec:lmxb} we carefully address the issue of contamination from low-mass X-ray binaries to the detected X-ray cores. \S~\ref{sec:res} and 5 presents our main results, specifically on the active fraction, dependence of accretion luminosity on black hole mass and X-ray luminosity function. We discuss the implications of our results in \S~\ref{sec:disc}, and conclude with summary in~\S~\ref{sec:conc}. 
We refer the reader to Paper I. for a thorough description of the program, as well as the determination of the parameters (such as stellar masses, stellar velocity dispersions, and black hole masses) employed throughout this series. In a companion paper, we report on the results from the \spi\ 24 $\mu$m observations of the same sample (Leipski \etal, Paper III.). 

\section{Data analysis}
\label{sec:data}
A detailed description of the \cxo\ data analysis, and the
cross-correlation to HST images is given in Paper I., \S\ 3.  Here we
summarize only the most relevant steps.  We observed each galaxy with
the Advanced CCD Imaging Spectrometer (ACIS) detector for 5.4 ks of
nominal exposure time in Faint-mode. The target was placed at the aim
point of the back-side illuminated S3 chip. Standard level 2 event
lists, processed for cosmic ray rejection and good time filtering,
were employed.  We first checked for background flares and removed
time intervals with background rate $\simgt 3\sigma$ above the mean
level.  Further analysis was restricted to energies between 0.3--7.0
keV in order to avoid calibration uncertainties at low energies and to
limit background contaminations at high energies.
We applied a wavelet detection algorithm over each activated chip,
using CIAO {\tt wavdetect} with sensitivity threshold corresponding to
a $10^{-6}$ chance of detecting one spurious source per PSF element if
the local background is uniformly distributed.
The \cxo\ astrometry was improved by cross-matching the detected
(non-nuclear) X-ray sources with the SDSS catalog (Data Release 5,
DR5), and the resulting bore-sight corrections were applied to the
data following the method described by Zhao \etal (2005).
Individual source locations are subject to statistical uncertainties
affecting the centroiding algorithm and to the dispersion of photons
due to the PSF. For ACIS-S, Garmire et al. (2000) estimate 90 per cent
confidences of $\pm$0\arcsec.5 for sources with $\sim$10 counts,
$\pm$0\arcsec.2 for 20-50 count sources, and negligible for $>$100
count sources. The statistical uncertainties affecting the centroid
errors in the positions of the X-ray sources, combined with the
$\simlt$0\arcsec.1 positional error of SDSS, results in a final
astrometric frame that is accurate to between 0\arcsec.2 (fields with
$\simgt$20 counts sources) and 0\arcsec.5 (fields with faint sources).
After registering the \cxo\ images to SDSS, we ran again {\tt
wavdetect} to refine the positions, and searched for point-like X-ray
emission centered at the galaxy optical center, derived from archival
\hst\ ACS images registered to the SDSS world coordinate system as
described in Paper I, \S\ 3.3.
We searched for X-ray counterparts to the optical nuclei within an error circle
which is the quadratic sum of the positional uncertainty for the X-ray source,
the uncertainty in the optical astrometry, and the
uncertainty in the X-ray bore-sight correction, multiplied by the chosen
confidence level scale factor (3$\sigma$).
All the newly 
detected X-ray nuclei are consistent with being point-like based on a comparison with the normalized point spread functions.
A circular region with a 2\arcsec\ radius centered on X-ray centroid
position, and an annulus with inner radius 20\arcsec\ and outer radius
30\arcsec\ were adopted for extracting the nuclear X-ray counts and for background subtraction, respectively. 
Given the low number statistics we are typically dealing with,  
we estimated the corresponding fluxes using
{\tt webPimms}\footnote{http://heasarc.gsfc.nasa.gov/Tools/w3pimms.html},
and assuming an absorbed power-law model with photon index $\Gamma=2$
and hydrogen equivalent column $N_{\rm H}$=$2.5\times 10^{20}$
cm$^{-2}$ (the nominal Galactic value determined from the HI studies of 
Dickey \& Lockman 1990). While the adopted photon index is relatively steep compared to the indices generally quoted in the literature for active nuclei ($1.4\simlt \Gamma \simlt 1.7$), this value is more representative of the low luminosity population ($1.6\simlt \Gamma\simlt 2$; e.g. Terashima \& Wilson 2003).  In particular, we obtained $\Gamma=2\pm0.2$ by stacking the detected X-ray nuclei in our sample and fitting their cumulative spectrum, as discussed in \S\ \ref{sec:stack}. 

Under the above-mentioned assumptions, $10^{-3}$ count \se\ in the 0.3-7 keV energy band
correspond to an intrinsic flux of 3.5$\times 10^{-15}$ \esc\ between 2-10
keV (ACIS-S). 
In case of no significant detection we applied Poisson statistics to derive
upper limits on the nuclear luminosity at the 95 per cent confidence level
(Gehrels 1986). Table~1. summarizes our results; 32/100 galaxies are found to host a X-ray core. While 51/100 show a massive nuclear star cluster (see Table 3 in Ferrarese \etal 2006b, for the fitted $g$ and $z$ magnitudes and half light radii), only 6 out of the 32 X-ray cores are classified as hybrid, i.e. hosting both a X-ray nucleus and a star cluster. 

\section{Low-mass X-ray binary contamination }
\label{sec:lmxb}

In a broad stellar mass range, and in the absence of a nuclear
star cluster, the total number of LMXBs and their cumulative X-ray
luminosity are proportional to the stellar mass of the host galaxy, \mstar\ (Gilfanov 2004; Kim \& Fabbiano 2004; Humphrey \& Buote 2004).  The number $n_X$ of expected sources per unit stellar mass
above a certain luminosity threshold can be estimated from the X-ray
luminosity function for LMXBs (the functional expression given by Gilfanov 2004 is employed throughout this series).  In turn, the number of
expected sources within the \cxo~PSF (convolved with the positional
uncertainty) is given by $n_X$ times $M_{\star, \rm PSF}$: the stellar
mass within the central aperture. As detailed in Paper I, the number of expected LMXBs
above the X-ray luminosity threshold of AMUSE-Virgo is typically lower than a few $10^{-2}$ at an average distance of
16.5 Mpc.  Given the functional expression of the LMXB luminosity function\footnote{Following Gilfanov (2004), the differential X-ray luminosity function of LMXBs in early types is parametrized as a power law with two breaks: $dn/dL_X $ scales as \lx$^{-\alpha}$, where $\alpha=1.0$ between $5\times 10^{35}-2\times 10^{37}$ erg \se, $\alpha=1.9$ between $2\times 10^{37}-5\times 10^{38}$ erg \se, and $\alpha=4.8$ above $5\times 10^{38}$ erg \se.}, it can also be ruled out that any
nuclear X-ray source be due to a collection of fainter LMXBs, as the
integral $\int dn/dL_X\cdot$ $d$\lx~is dominated by the luminosity
term.

As discussed in the introduction, the very existence of SMBHs in faint inactive galaxies remains questionable. In fact, Ferrarese
\etal (2006a) suggest that compact stellar nuclei tend to be more common than nuclear SMBHs at the low-mass/low-luminosity end of the galaxy mass function, possibly taking over at magnitudes fainter than $-18$ (this does not necessarily implies that nuclear SMBHs and star clusters can not possibly coexist; see e.g. Seth \etal 2008; Graham \& Spitler 2009).
For the purpose of this work, 
the presence of a nuclear star cluster, having an enhanced stellar
encounter rate, implies a higher chance contamination from LMXBs with
respect to the nuclei of massive early type galaxies, which, on the contrary, show a light deficiency with respect to standard Sersic profiles (Ferrarese \etal 2006a; C\^ot\'e \etal 2006, 2007; Kormendy \etal 2009).  
In order to
account for this effect, in the presence of a nucleated galaxy we
conservatively adopt the X-ray luminosity function of LMXBs in
{\it globular clusters}, as estimated by Sivakoff \etal (2007).  It is
known that, while hosting a small percentage of the galaxy stellar
mass, globular clusters are home to about 50$\%$ (with an admittedly large scatter) of the observed
LMXBs. In this environment, the number of expected LMXB sources scales
non-linearly with the cluster mass.  Sivakoff \etal (2007) derive an
expression for the expected number\footnote{$ n_X \propto 8\times 10^{-2}
\big({M_s}/{10^6 M_{\odot}}\big)^{1.237} \big({r_{\rm h}}/{1~{\rm
pc}}\big)^{-2.22} $\\.} $n_X$ of LMXBs brighter than
$3.2\times 10^{38}$ erg \se\ (the luminosity limit is set by the
sample completeness) in a star cluster of stellar mass $M_s$, half
mass radius $r_{\rm h}$ (incidentally, this is about the same completeness limit of our Virgo sample).

For our faint, nucleated spheroidals, the mass of the nuclear star cluster
can be estimated from Equation 1 in Ferrarese \etal (2006a), using the colors and half-light radii given in Table 3 by Ferrarese \etal (2006b). The expected number of 
LMXBs in a globular cluster can then be converted to a probability $P_X$
that there is at least one LMXB brighter than the detected X-ray
luminosity assuming a Poisson distribution. 
In terms of chance probability of hosting a X-ray active SMBH, each of the 6 galaxies which host both a nuclear X-ray sources and a massive star cluster is then assigned a weight  ($1-P_X$): see Table~2. for details\footnote{As discussed in Paper I, the optical radial profile of VCC1178 also shows marginal evidence for a nuclear star cluster. We estimate the probability of hosting a LMXB within the \cxo\ PSF to be $\sim 12\%$, hence $w=1-P_X$=0.88.  However, we assigned $w=1$ to the nuclear X-ray source in VCC1178 based on independent evidence that this galaxy hosts a nuclear SMBH, specifically the presence of a double optical nucleus  (see Lauer \etal 1996).}. This yields our final, {\it weighted} distribution of galaxies hosting a active (down to our luminosity threshold) SMBH.

\subsection{Stacking Analysis}
\label{sec:stack}

\begin{figure}
\vspace{4cm}
\center{\includegraphics[angle=0,scale=.36]{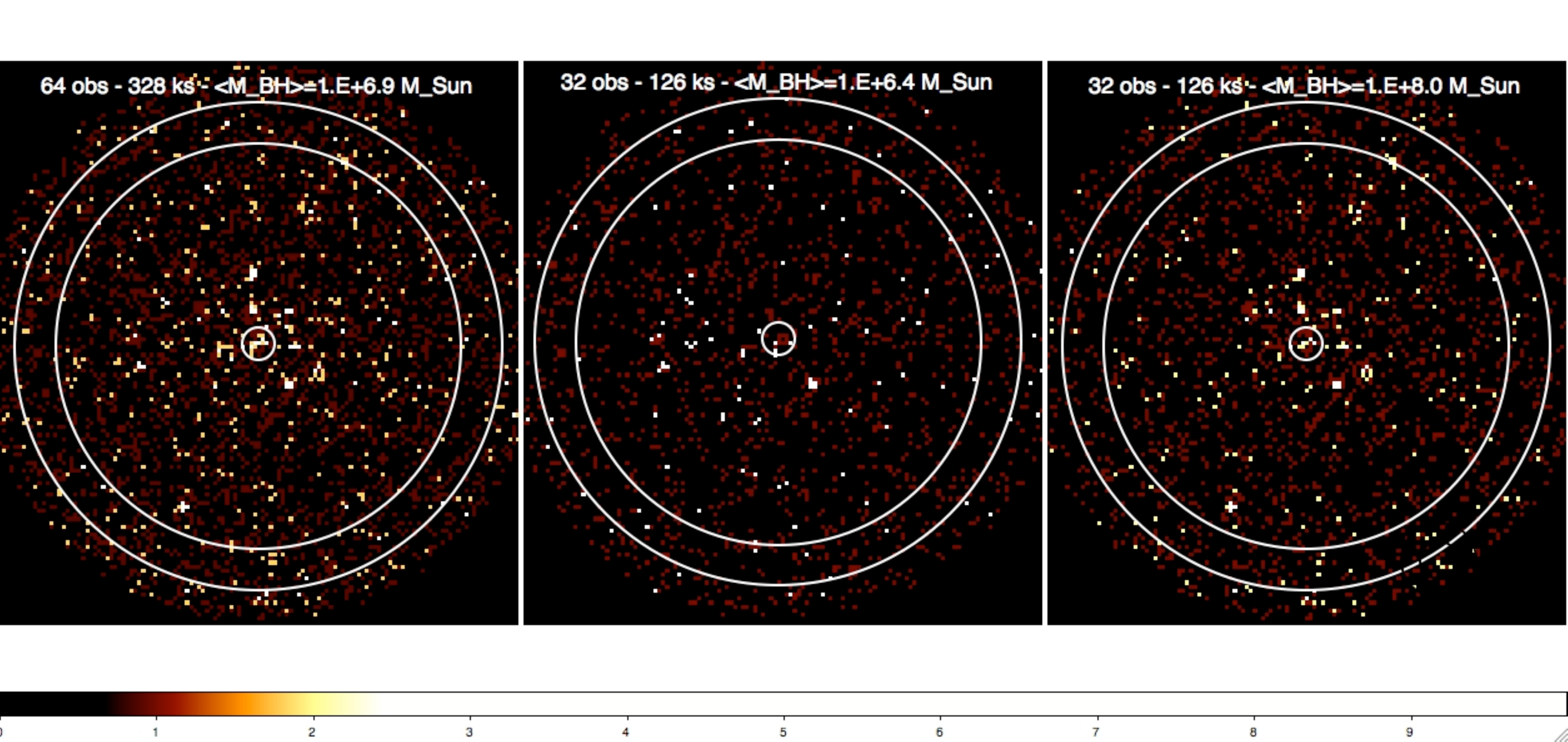}
\vspace{1cm}
\caption{\label{fig:stack}Stacking undetected nuclei. {\it Left}: The inner 35\arcsec\ apertures of 64 AMUSE-Virgo galaxies with undetected X-ray cores have been stacked to give a single, 328 ks ACIS-S image. A nuclear X-ray source is significantly detected in the stacked image, with net count rate of $1.1\pm0.2\times 10^{-4}$ count \se. 
The corresponding average X-ray luminosity is consistent with what expected from nuclear LMXBs within the \cxo\ PSF.
}}
\end{figure}

In order to place more stringent limits on the average flux, we stacked the 0.3-7 keV images of the 64 non-detected nuclei with snapshot observations\footnote{For the stacking analysis, we only used those galaxies with snapshot, 5.4 ks observations, for a total of 64 targets. The remaining 4 galaxies with non-detections are: VCC1226, VCC0881, VCC798 and VCC1535. Being massive ellipticals with deep archival exposures, all four galaxies have substantial contamination to the soft X-ray band from diffuse gas emission, which is the main reason why we did not include them in the stacking analysis. In these 4 cases, the search for nuclear X-ray sources was performed in the hard X-ray band only, as described in detail in \S\ 3.2 of Paper I.}, for a total of 328 ks of effective exposure.  We extracted the counts from a 2\arcsec\ radius circular aperture,
and background from an annulus with inner and outer radii $R_{\rm
in}=$25\arcsec\ and $R_{\rm out}$=30\arcsec, centered on the stacked
nuclear position (see Figure~\ref{fig:stack}, left panel).  49 counts are detected within the
2\arcsec\ radius aperture, while 12.8 are expected from the background, indicating a statistically significant detection (null detection probability lower than $2\times 10^{-7}$). 
At a distance of 16.5 Mpc, and assuming an absorbed power law model as discussed in the previous section, the measured net count rate ($1.0\pm0.2\times 10^{-4}$ count \se) corresponds to an {\it average}, unabsorbed 0.3-7 keV luminosity $\langle$\lx$\rangle=3.5\pm 0.7 \times 10^{37}$ erg \se.
This emission can be accounted for by an ensemble of low-luminosity LMXBs within the total stellar mass enclosed by the \cxo~PSFs: following Gilfanov (2004), the expected X-ray luminosity due to LMXBs within all of the 64 nuclei (enclosing about $10^{10}$ \msun) is $6\pm2 \times 10^{37}$ \es. 
In terms of average Eddington scaled luminosity, this sets an upper limit of $3\times 10^{-8}$ for an average nuclear black hole mass $\langle$\bhm$\rangle$=$10^{6.9}$ \msun. 

We also grouped the 64 observations into two sets of 32 observations corresponding to two black hole mass bins, with $\langle$\bhm$\rangle_a = 10^{8.0}$ \msun\ and $\langle$\bhm$\rangle_b = 10^{6.4}$ \msun, and stacked them into 2 observations by 164 ks. Both show a statistically significant nuclear detection: the measured net count rates for observations $a$ and $b$ are $1.3\pm0.3\times 10^{-4}$ count \se, and $0.7\pm0.2\times 10^{-4}$ count \se. Again, the inferred luminosities are consistent with what expected from faint nuclear LMXBs. They translate into the following Eddington-scaled luminosities: $\langle$\lx/\ledd$\rangle_a$$<3.0\times 10^{-9}$ and  $\langle$\lx/\ledd$\rangle_b$$<6.4\times 10^{-8}$, respectively. The middle and right panels of Figure~\ref{fig:stack} show the two stacked images with 32 galaxies each: low-\bhm\ bin and high-\bhm/ bin, respectively.

Finally, in order to characterize the average spectral properties of our sample, we stacked the images of all the detected X-ray nuclei with snapshot observations, for a total of 102 ks of effective exposure; $236\pm16$ counts are detected from stacked nuclear X-ray source, corresponding to a net count rate of $2.3\pm0.1\times 10^{-3}$ count \se (0.3-7 keV).  
We constructed the soft (S), medium (M) and hard (H) band stacked images, restricting the energies between 0.3-0.7 keV, 0.7-1.1 keV, and 1.1-6 keV, respectively (e.g. Ott, Walter \& Brinks 2005). The net counts in each band  (S$=30.1\pm5.9$, M$=73.1\pm8.8$, and H$=130.8\pm8.0$) yield the following hardness ratios for the stacked nucleus: HR1=(S-M-H)$/$(S+M+H)$=-0.7\pm0.1$, and HR2=(S+M-H)$/$(S+M+H)$=-0.1\pm0.1$. These relatively hard values of HR1 and HR2 are quite typical of accretion powered sources, such as AGN or X-ray binaries, confirming our interpretation of low-level accretion-powered emission (rather than thermal emission from hot gas, as observed e.g. in star forming regions). 

\section{The duty cycle of nuclear activity} 
\label{sec:res}

\subsection{The active fraction vs. host stellar mass}
\label{ssec:af}

The {\it weighted} (as described in \S~\ref{sec:lmxb}) distribution of galaxies hosting a X-ray active SMBH down to our luminosity limit  is plotted in Figure~\ref{fig:activefrac} as a function of the host stellar mass \mstar.
Overall, we infer that between $24-34\%$ of the
galaxies targeted by AMUSE-Virgo host a X-ray active SMBH, down to a luminosity threshold of $\sim 2\times 10^{38}$ \es\ (2-10 keV, at the 95$\%$
C.L.). {\it This sets a firm lower limit to the black hole occupation
fraction in nearby bulges within a cluster environment}.

Perhaps not surprisingly, the incidence of nuclear activity increases with stellar mass: between $0.7-14\%$ of the galaxies with log(\mstar)~between $8.5-9.5$ host a nuclear, active SMBH; the fraction raises to $16-43\%$ for $9.5<$log(\mstar)$<10.5$, up to $53-87\%$ for  log(\mstar)$>10.5$ (at the 95$\%$ confidence level). 
The known trend (since the first Palomar sample; Ho \etal 1997) of increasing active SMBH fraction with increasing host stellar mass is obviously amplified by (possibly entirely driven by) the fact that, going down the mass function, as the nuclear BH masses decrease with (some power of) \mstar, so do their expected X-ray luminosities for a fixed Eddington ratio. 

In order to assess this completenss effect, we estimated the fraction of X-ray active SMBHs as a function of \mstar~ for four sub-samples, each complete down to \lx/\ledd$=-9,-8,-7,-6$. These `Eddington-complete' active fractions are listed in Table~\ref{tab:completeness} as a function of \mstar: 
within the limitations of the admittedly large error bars imposed by such low statistics sub-samples, this exercise proves that {\it once the issue of Eddington completeness is taken into account, there is no evidence for a statistically significant increase in the incidence of nuclear SMBH activity with increasing host stellar mass}.  
For reference, Figure~\ref{fig:comparison} illustrates the distribution of the inferred Eddington scaled X-ray luminosities for the AMUSE-Virgo sample as a function of \bhm~(in red), and compares it with previous works (Pellegrini \etal 2005; Soria \etal 2006a,b; Zhang \etal 2009). 
While the distribution of active SMBHs is quite broad in terms of Eddington-scaled X-ray luminosities, with \ledd\ ranging between a few $10^{-9}$ up to a few $10^{-6}$, this plot seems to show a slight increase of the Eddington-scaled X-ray luminosity with black hole mass. We quantify this in the following section.

\subsection{Accretion luminosity vs. black hole mass}
\label{ssec:likeli}

We have chosen to focus on the Eddington-scaled X-ray luminosity as the more
physically meaningful quantity; to investigate more rigorously the connection between accretion-powered X-ray 
luminosity and black hole mass, in this section we infer the parameters of a power law relationship between the measured \lx\ and
\bhm\ through a Bayesian approach:
\be
\log (\hat{L}_{X}/10^{38}{\rm erg\ s^{-1}})  = 
   A + B \ \log (\hat{M}_{\rm BH}/10^8 M_{\odot})
\label{eq:pjm}
\ee
where the hats indicate true, underlying quantities -- as opposed to the
noisy, observable quantities \lx\ and \bhm.

This is equivalent to considering the dependence of the {\it average} Eddington-scaled X-ray luminosity with black hole mass: $\langle$\lx/\bhm$\rangle\propto\langle$\lx/\ledd$\rangle\propto$ \bhm$^\beta$, where $\beta=(1-B)$. 
We assume errors $\sigma_M$=0.44 dex on the logarithm of the black hole mass (G\"ultekin \etal 2009),
30\% uncertainty on the measured X-ray luminosities, and use this
to define Gaussian likelihood functions ${\rm , {P_r}} ( M_{\rm BH}|\hat{M}_{\rm
BH})$ and ${\rm {P_r}} ( L_{\rm X}|\hat{L}_{\rm X})$. For the non-detections, 
${\rm {P_r}} ( L_{\rm X}|\hat{L}_{\rm X})$ is taken to be Gaussian for
luminosities higher than our limiting threshold, and uniform for all
luminosities less than the limit. 
 
We include intrinsic scatter~$\sigma_0$ on the luminosity as a free parameter
as well, via a Gaussian probability density function (PDF) ${\rm P_r}
(\hat{L}_{\rm X}|\hat{M}_{\rm BH})$.  We assign scale-invariant Jeffreys priors
both for this parameter and also the power law normalization (see e.g. Sivia
2006).  For the power law index $B$, we note that this is a gradient on a
log-log plot, and, in the absence of any causal connection between \lx\ and
\bhm, we assign a prior to enforce rotational invariance, which is equivalent
to retaining ambivalence as to whether we are seeing an \lx-\bhm\ relation, or
rather a \bhm-\lx\ relation. 

This results in the following PDF:
\be
{\rm {P}_r}(B) \propto \left( 1 + B^2 \right)^{-3/2}.
\ee

Having defined these distributions, we then, at any given position in \{$A,B,
\sigma$\} space, integrate out each of the $i=100$ nuisance parameters
$\hat{M}_{{\rm BH},i}$. This operation is equivalent to convolving the
intrinsic spread with the error distributions: 

\be
{\rm P_r}({L}_{{\rm X},i}|{M}_{{\rm BH},i},A,B,\sigma_0){} = 
   \frac{\exp { \displaystyle\left[{ -\frac{ (\log{L}_{\rm X} - B \log{M}_{\rm BH} -A)^2 }
                                           {2(\sigma_0^2 + \sigma_{{L},i}^2 + B^2 \sigma_{{M},i}^2)}}
                \displaystyle\right]}} 
        {\sqrt{(2 \pi)(\sigma_0^2 + \sigma_{{L},i}^2  + B^2\sigma_{{M},i}^2)}}\,
\label{eq:convlhood}
\ee
where $\sigma_L$ denotes the error on $\log$\lx. This marginalization requires a
prior to be assigned for the $\hat{M_{\rm BH}}$ parameters: 
we used a uniform distribution
in $\log(\hat{M_{\rm BH}})$, which corresponds to a power-law mass function of
index~$-1$ (see e.g. Greene \& Ho 2007, 2009 for comparison).

We explore the \{$A,B, \sigma$\} parameter space using a Markov Chain Monte-Carlo
sampler, the output of which is a set of samples that characterize the desired
posterior PDF ${\rm P_r} (A,B,\sigma|\{L_{\rm X},M_{\rm BH}\})$.  We ran
multiple chains to ensure convergence. The fit is illustrated in
Figure~\ref{fig:line}, where we show the (median) intrinsic scatter about the
(median) relation, while the posterior distribution is shown in
Figure~\ref{fig:pdf}, where we have marginalized over the normalization~$A$. 
We find the index~$B = 0.38^{+0.13}_{-0.12}$, while the scatter is inferred
to be $\sigma_0=0.46^{+0.08}_{-0.06}$~dex (68\% confidence intervals around the
one-dimensional marginalised PDF median values). 

A zero slope, corresponding to no dependence of \lx\ on
\bhm, is ruled out at high confidence: ${\rm P}_r(B = 0.0) \simeq 0.001$.
Likewise, a slope of unity, corresponding to no dependence of \ledd\ on \bhm, is also rejected: ${\rm P}_r(B = 1.0) \lesssim
10^{-4}$.
In terms of Eddington ratio, the relation in Equation~\ref{eq:pjm} still becomes:
$\langle$\lx/\ledd$\rangle\propto$ \bhm$^{-0.62}$, that is, on average, {\it low mass black holes seem to
be more active (i.e. emit closer to their Eddington limits) than higher mass
objects}. As a consequence --pending the large uncertainties on the black holes' mass function (see Greene \& Ho 2007, 2009) and occupation fraction (see Volonteri \etal 2008; Van Wassenhoven \etal 2010) in the local universe-- the active fraction, here defined as above a certain Eddington ratio, is a decreasing function of \bhm. 

As how to the average Eddington-scaled X-ray luminosity (and hence the active fraction) might depend on \mstar, this is obviously related to the scaling of \bhm\ with \mstar\ (such a scaling is somewhat expected, albeit with a much larger scatter, as a result of the observed relation between \bhm\ and bulge mass). Specifically, for the black holes in our sample of active SMBHs, \bhm$ \propto $\mstar$^{\delta}$, with $\delta=1.2\pm0.6$ (assuming a scatter of 0.44 dex for both). Owing to the extremely large scatter, this could still be marginally consistent with no dependence of \ledd\ (hence of the active fraction) on stellar mass, \ie: $\langle$\lx/\ledd$\rangle \propto$ \mstar$^{-0.8\pm0.5}$ ($68\%$ C.L.). 
\section{The X-ray luminosity function}
\label{sec:xlf}

The differential logarithmic 
X-ray luminosity function (XLF) of the active SMBHs in the AMUSE-Virgo sample is well approximated by the following expression:
\begin{equation}
\frac{dN}{d{\rm ln}L_{\rm X}} = K_{_{\rm \small AMUSE}} \left(\frac{L_{\rm X}}{10^{38} {\rm erg~s^{-1}}}\right)^{-\Gamma_{_{\rm \small AMUSE}}}
\end{equation}
where the best-fit parameters are: 
$K_{\rm AMUSE}=1.4\pm0.1$, and $\Gamma_{\rm AMUSE}=0.4\pm0.1$. 
For comparison, the differential XLF of off-nuclear LMXBs, as given by Gilfanov (2004), is well approximated by a power law with two breaks, having negative differential slopes\footnote{Defined as $\beta=(\Gamma+1)$.}  
$\beta_1=1.86\pm 0.12$ between $0.2-5.0\times 10^{38}$ erg \se, and $\beta_2=4.8\pm1.1$ above $5\times 10^{38}$ erg \se, while its normalization scales with the stellar mass of the host galaxy (see also: Kim \& Fabbiano 2004; Hamphrey \& Buote 2004).  
A visual comparison between the two XLFs is also shown in Figure~\ref{fig:xlf}: plotted are the fitted XLF for the nuclear X-ray sources in the AMUSE sample (red histogram) and the functional expression for the LMXB XLF (Gilfanov 2004), scaled to a stellar mass of $10^{11}$ \msun. On average, this normalization corresponds to conservatively assuming that the \cxo\ PSF at the distance of Virgo encloses about $2\%$ of the stellar mass of each galaxy. 
The fitted XLF slope is much shallower for the AMUSE sample than for the extragalactic LMXBs, further strengthening our conclusion that contamination from stellar mass objects has been properly accounted for. In addition, we recover a similar slope to that reported by Zhang \etal (2009) for a sample of 86 nearby (within 15 Mpc) galaxies hosting a nuclear X-ray source: the fitted slope for their sample is 0.5, i.e. consistent, within the errors, with our results. We argue though that the Zhang \etal sample might suffer from a slight contamination from X-ray binaries at the low mass end, given that it includes both early and late type galaxies -- with the latter ones having an enhanced contamination from bright high-mass X-ray binaries.

\section{Discussion}
\label{sec:disc}

\subsection{Super-massive black holes vs. massive nuclear star clusters}

Until recently it was generally believed that massive black holes and
nuclear star clusters did not generally coexist at the centres of galaxies. 
Less than a handful of counter-examples (e.g. Filippenko \& Ho 2003; Graham \& Driver 2007) were the exceptions to confirm the rule. 
More systematic studies, e.g. by Ferrarese \etal (2006a) and Wehner \& Harris (2006), showed the transition between galaxies which host predominately a black hole vs. a nuclear 
cluster occurs around 10$^{10}$ \msun. 
However, while the latter conclude that nucleated galaxies show no evidence of hosting SMBHs, the former speculate that nuclei form in all galaxies but they are destroyed by the evolution of preexisting SMBHs or collapse into a SMBH in the most massive cases. It is also suggested that ``SMBHs and nuclei are almost certainly mutually exclusive in the faintest galaxies belonging to the VCS sample\footnote{The same considered in this work.}". This is indicated by the fact that, although the nuclear masses of NGC~205 and M33 are fully consistent with the relation\footnote{See Figure 2 in Ferrarese \etal (2006a).} between the mass the nuclear object (be it a cluster or a BH) and the host galaxy, the upper limits on their SMBH masses are not, implying that {`` [..] neither galaxy contains an SMBH of the sort expected from extrapolations of the scaling relations defined by SMBHs in massive galaxies"}. 
Merritt (2009) examines the evolution of nuclear star clusters with and without SMBHs from a theoretical point of view, finding that nuclear star clusters with black holes are always bound to expand, due primarily to heating from the galaxy and secondarily to heating from stellar disruptions. As a consequence, core-collapsed clusters should not be harboring nuclear black holes. 

From an observational point of view, there has been a number of efforts to quantify the degree of coexistence of nuclear clusters and SMBHs, and their mutual properties. 
Seth et al.\ (2008) searched for active nuclei in 176 galaxies with known nuclear clusters, using optical spectroscopy, X-ray and radio data. They find that the AGN fraction increases strongly with increasing galaxy and nuclear cluster mass, consistent with previous studies of the general galaxy population.  In addition, the variation of the AGN fraction with Hubble type is also consistent with the whole Palomar sample (Ho \etal 1997), indicating that the presence (or absence) of a nuclear star cluster does not play a crucial role in boosting (or hampering) accretion-powered activity onto a SMBH (see also Gonzalez-Delgado et al.\ 2008).
More recently, Graham \& Spitler (2009) reported on 12 new systems which host both a nuclear star cluster and a SMBHs, and for which they were able 
acquire both the masses of the nuclear components, as well as the stellar mass of the host spheroid. They find that, for host stellar masses in the range  $10^{8-11}$ \msun, the nucleus-to-spheroid mass ratio decreases from a few to about 0.3$\%$. This ratio is expected to saturate to a constant value once dry merging commences, and the nuclear cluster disappear. 
Our work tackles the issue of nuclear SMBH-star cluster coexistence taking a complementary approach with respect to that of Seth et al., that is, we ask how many of those nuclei which show a nuclear cluster also host a (X-ray active) SMBH. 
We remind the reader that the VCS sample (C\^ot\'e \etal 2004) surveyed by AMUSE-Virgo is complete down to a $B$ magnitude of $-$18, and is a random sample of fainter (early type) objects. In terms of (hosts') stellar mass distribution, the sample peaks well below $10^{11}$ \msun, and is thus particularly well suited for investigating hybrid potentially nuclei. 
While 32 out 100 galaxies are found to host a nuclear X-ray source, only 6 of them also show evidence for a nuclear star cluster as visible from archival \hst~images (typically, the star clusters are identified as overdr-densities above a single-component Sersic profile, but see Ferrarese \etal 2006b for details about the fitting procedure). After taking into account LMXB contamination, while the fraction of hybrid nuclei as function of host stellar mass \mstar\ is constrained between 0.3 and 7$\%$ for log\mstar$>11$ (95$\%$ C.L., and down to a limiting 2-10 keV luminosity of $\sim 2\times 10^{38}$ \es), the lack of star cluster--SMBH matches above $10^{11}$ sets an upper limit of 32$\%$ to the fraction of such hybrid nuclei in massive early types.

\subsection{Active fraction and down-sizing in black hole accretion}

The bottom panel of Figure~\ref{fig:activefrac} shows how he fraction $f_{X}$ of objects hosting an active SMBH --down to our luminosity limit-- increases as a function of the host mass \mstar: $0.01<f_X<0.14$ for log(\mstar/\msun)$<{9.5}$, $0.53<f_X<0.87$ for log(\mstar/\msun)$>{10.5}$, and $0.16<f_X<0.43$ for intermediate masses.
While this result is fully consistent with earlier works (Ho \etal 1997; Decarli \etal 2007; Kauffmann \etal 2003; Seth \etal 2008; Satyapal \etal 2008; Gallo \etal 2008), {this is not to say that the fraction of objects which host an active SMBH raises with mass tout-court}.
As discussed in \S~\ref{ssec:af}, dealing with a sample that is `Eddington-limited', rather than simply luminosity-limited, results in {no observational evidence for a statistically significant increase in the active SMBH fraction with mass (either host galaxy's or BH's)}. 
The same test as above can be applied to hybrid nuclei: after considering sub-samples complete down to Eddington ratios between $10^{-6}$ and $10^{-9}$, we conclude that the number of nuclei hosting both an active SMBH and a massive star cluster does not show a trend of increasing incidence with increasing host stellar mass. It must be stressed, however, that restricting the analysis to `Eddington-complete' sub-samples results in very large error bars. 

A more quantitative constraint can be obtained through a Bayesian approach to infer a dependence between the measured X-ray luminosities and black hole masses (\ref{ssec:likeli}).
{This analysis highlights, for the first time, a dependence of accretion-powered X-ray luminosity on black hole mass. A slope $B=0.38^{+0.13}_{-0.12}$ in Equation 1 implies that the average Eddington scaled X-ray luminosity scales with black hole mass to the power $-0.62$. 
As a consequence, {\it the local active fraction}-- defined as
those above a fixed X-ray Eddington ratio-- {\it decreases with black hole mass}} (as well as with host stellar mass, as long as a positive relation holds between those two quantities). 
{This can be considered as the analogous of galaxy `down-sizing' (cf. Cowie \etal 1996) in the context of black hole accretion: within the local universe,  low mass SMBHs emit closer to their Eddington limit than high mass objects.} 
On the same line, using a combination of SDSS and Galaxy Zoo data within $z=0.05$, Schawinski \etal (2010) find evidence for a dependence of the black hole growth on the host galaxy morphology: while late-type galaxies host preferentially the most massive black holes, early-type galaxies host preferentially low mass black holes. More specifically, for (bolometric) Eddington-scaled luminosities in excess of 10$\%$, low mass early type galaxies are the only population to host a substantial fraction of active SMBHs, in qualitative agreement with our findings (albeit at higher luminosities).\\

A second interesting result is that, within the VCS sample surveyed by AMUSE-Virgo, the distribution of Eddington-scaled luminosities of the detected nuclei turns out to be very broad, in contrast with what reported for higher redshift {\it bona fide} AGN. 
In fact, Figure~\ref{fig:comparison} indicates no clear cut in terms of active SMBH luminosity: as stated at the beginning of this Paper, 
the distinction between active and inactive is rather arbitrary, and is ultimately set by our ability to detect and interpret signatures of accretion-powered activity. 
In principle,  X-rays due to non-thermal processes in the vicinity of a BH offer clear-cut diagnostics of accretion-powered activity. 
However, as illustrated in Table~2. already at the distance of Virgo, the chance contamination to the nuclear X-ray emission from LMXBs is substantial, even with the fine spatial resolution of \cxo. X-ray surveys of `inactive' galaxies are necessarily limited by this contamination; the problem is only marginally alleviated with deeper exposures, since the 0.5-10 keV spectrum of a luminous X-ray binary is virtually undistinguishable from that of a highly sub-Eddington SMBHs.  As the distance increases so does the mass enclosed with the X-ray instrument PSF, and thus the chance of having a number of medium luminosity X-ray binaries `mimicking' a low luminosity AGN. While X-rays alone can not possibly provide us with a clean answer as to the nature of the nuclear accretors, the picture may clarify when the high energy properties are interpreted in concert with other wavelength's. In particular, mid-infrared emission can be a sensitive tracer of accretion-powered emission from a massive accretor: while X-ray binaries emit the bulk of their energy in the X-ray band, SMBHs accretion-powered emission ought to peak at lower frequencies, thus representing a potential source of heating for local dust to re-process. Additionally, sensitive mid-IR observation may unveil obscured accretion-powered emission due to the presence of e.g. a dust lane/dusty torus. 
This will be explored in detail by comparing the mid-IR (24 $\mu$m \spi\ observations) radial profiles of the AMUSE-Virgo galaxies with the smoothed optical profiles from \hst\ (Leipski \etal, Paper III., in preparation).\\

The notion that energy
feedback from super-massive black holes could solve a number of problems faced
by the hierarchical paradigm at galactic scales has been substantiated by a number of recent works: the vast majority of AGN at $z\simeq 1$ occupy a rather distinct region of color-magnitude diagram, typically associated with the end of the star-forming phase (the so called "green valley"; e.g. Schawinski \etal 2007, 2009; Salim et al. 2007; Nandra et al. 2007; Silverman et al. 2008; Bundy et al. 2008; Georgakakis et al. 2008).
It is also well known that environment plays an important role in
regulating the gas supply in galaxies, and as a consequence their star
formation rate and morphologies, via a variety of processes including
high speed interactions, mergers, ram-pressure stripping. It has often been proposed that
galaxy mergers and interactions also regulate fueling of the central
SMBH.  Although the situation remains
controversial, progress is being made regarding the role of
environment in regulating activity at high Eddington ratios. For
example, observations of large samples of optically-selected AGN from
SDSS show evidence that the luminous optically active AGN are
associated with young stellar populations (Kauffmann et al. 2003; Choi \etal 2009). 
At the extreme end, this scenario is
supported by observations of ultra-luminous infrared galaxies, that
are in general associated with galaxy mergers, and have bolometric
luminosities similar to quasars. High resolution numerical simulations
(Springel, Di Matteo \& Hernquist 2005; Ciotti \& Ostriker 2007) show that the AGN activity remains obscured during most of the
starburst and AGN activity phase. Sub-millimeter observations of samples of absorbed vs. unabsorbed AGN have consistently confirmed this picture (e.g. Page \etal 2004; see also Alexander \etal 2008).

In contrast to this, low-luminosity
radio-loud AGN in the nearby universe are seen to be preferentially
hosted by massive elliptical galaxies, which tend to be found in
richer environments, where gas-rich galaxy mergers are less likely to
occur.  Comprehensive studies are underway to characterize the AGN
populations in clusters vs field as a function of cosmic time.
However, very little in is known about the environmental dependency of
nuclear activity at low levels. Given the suggested role of low levels
of nuclear activity in regulating star formation, the discovery of environmental trends would have profound
implications not only for understanding black hole growth but also for
understanding galaxy formation and evolution.
This issue will be tackled through an approved \cxo~ program (Cycle 11) targeting 100 early type field galaxies within 30 Mpc. \\

\section{AMUSE-Virgo. II: Summary}
\label{sec:conc}

\begin{figure}
\center{\includegraphics[angle=0,scale=.7]{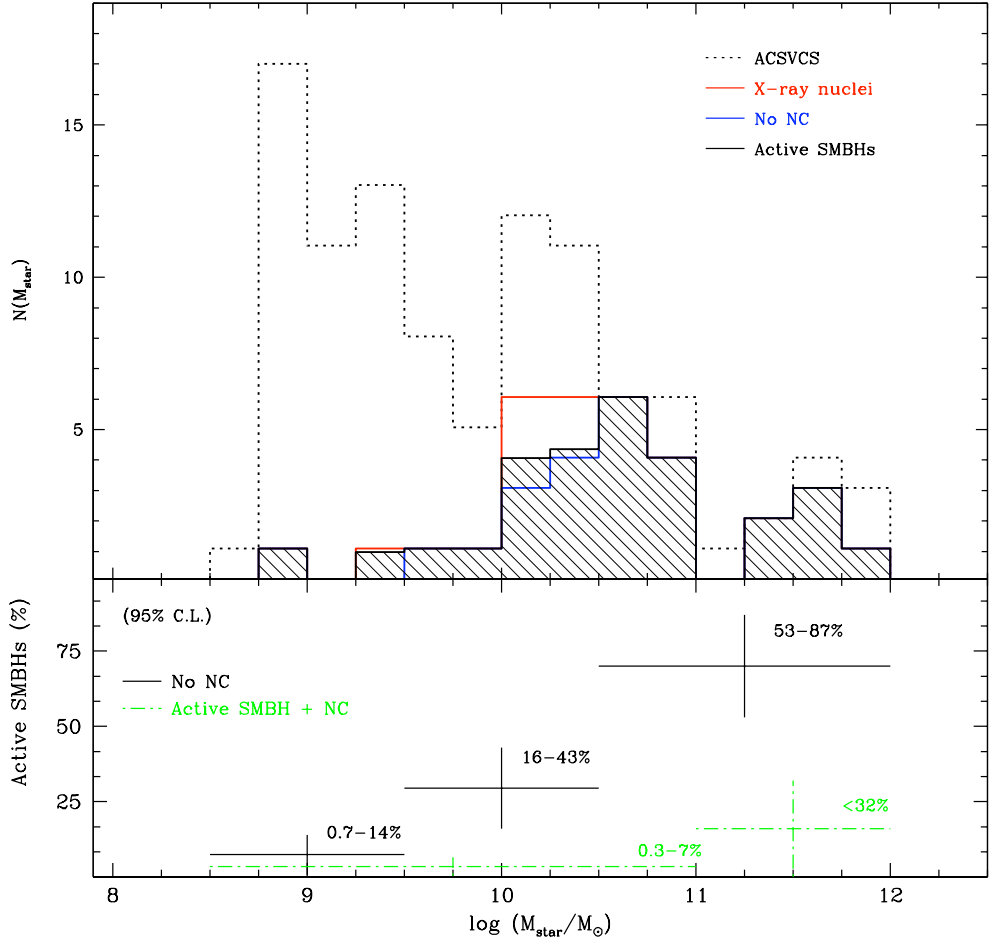}
\vspace{-3cm}
\caption{{The top panel illustrates the distribution of the 100 galaxies targeted by AMUSE-Virgo (the ACSVCS sample) as a function of the host stellar mass: The dotted histogram represents the whole sample; the red histogram is for the 32 galaxies found to host X-ray cores, while the blue histogram is for the 26 X-ray cores that host no nuclear star cluster. The 6 galaxies which host hybrid nuclei have a higher chance contamination to the nuclear X-ray emission from LMXBs. The `weighted' distribution of galaxies hosting an accreting SMBH is illustrated by the black shaded histogram (see \S~\ref{sec:lmxb}). The bottom panel shows the active SMBH fraction (solid black line), and the fraction of hybrids (dot-dashed green line) as a function of the host stellar mass. Numbers at given at the 95$\%$ confidence level.
\label{fig:activefrac}}}}
\end{figure}

\begin{figure}
\center{\includegraphics[angle=0,scale=.65]{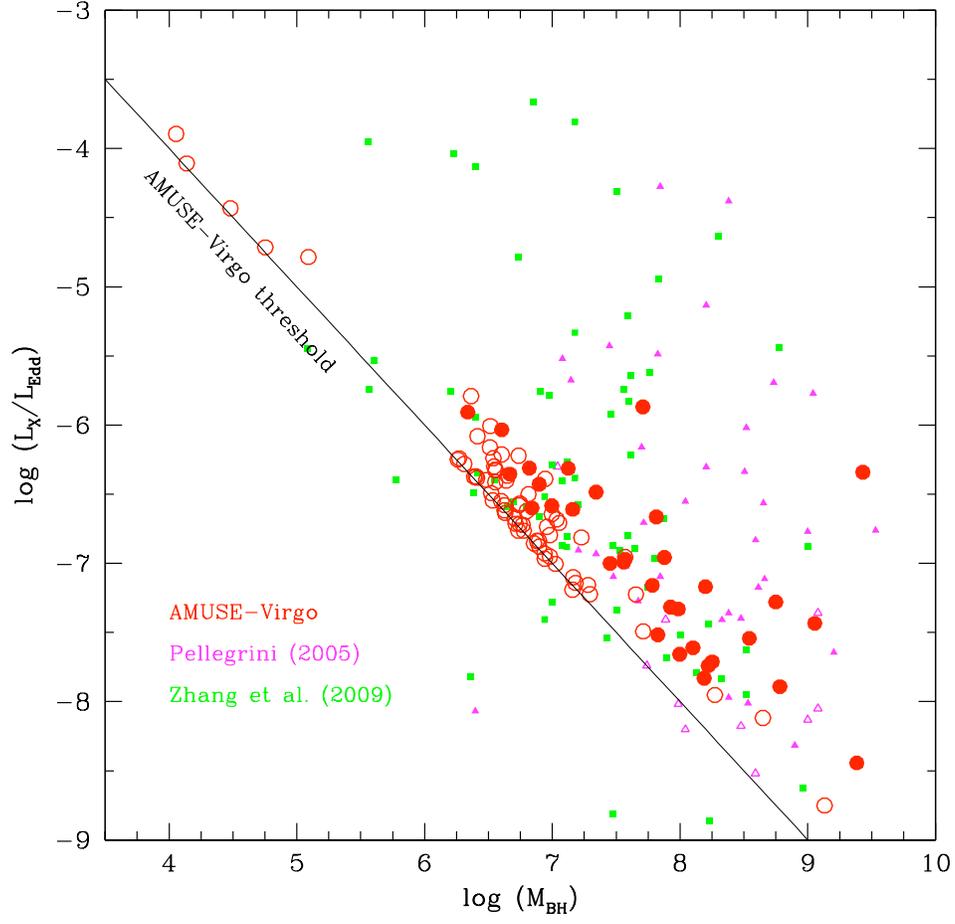}
\vspace{-2cm}
\caption{The results from AMUSE-Virgo, in red, are compared to those presented by Pellegrini (2005) and Zhang \etal (2009), in magenta and green, respectively. Open marks represents upper limits, while filled marks are for detections.  Undetected nuclei from AMUSE-Virgo (open red circles) which are not `collapsed' onto the \cxo~detection threshold for our survey indicate only marginal detections (less than 6 photons were detected within a radius of 2\arcsec\ centered on the optical position).  \label{fig:comparison}}}
\end{figure}

\begin{figure}
\vspace{2cm}
\center{\includegraphics[angle=0,scale=.65]{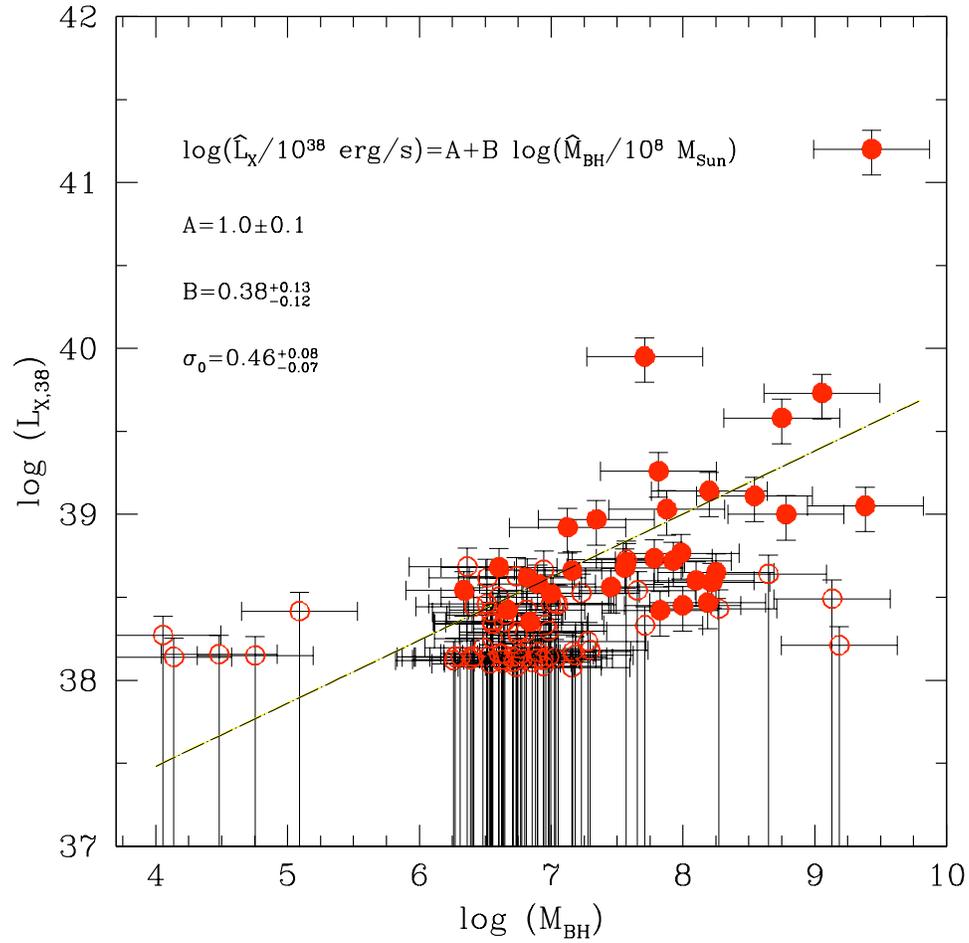}
\vspace{-2cm}
\caption{{Inferred X-ray luminosities/limits for the 100 AMUSE Virgo galaxies, plotted as a function of the nuclear black hole mass. The maximum likelihood analysis presented in \S\ \ref{ssec:likeli} shows evidence for a correlation of the form: $\langle$\lx$\rangle\propto\langle$\bhm$\rangle ^{0.38}$, implying that the average Eddington scaled X-ray luminosity scales with \bhm\ to the power $-0.62$. This argues for a {`down-sizing' in black hole accretion}.
\label{fig:line}}}}
\end{figure}

\begin{figure}
\vspace{2cm}
\center{\includegraphics[angle=0,scale=.55]{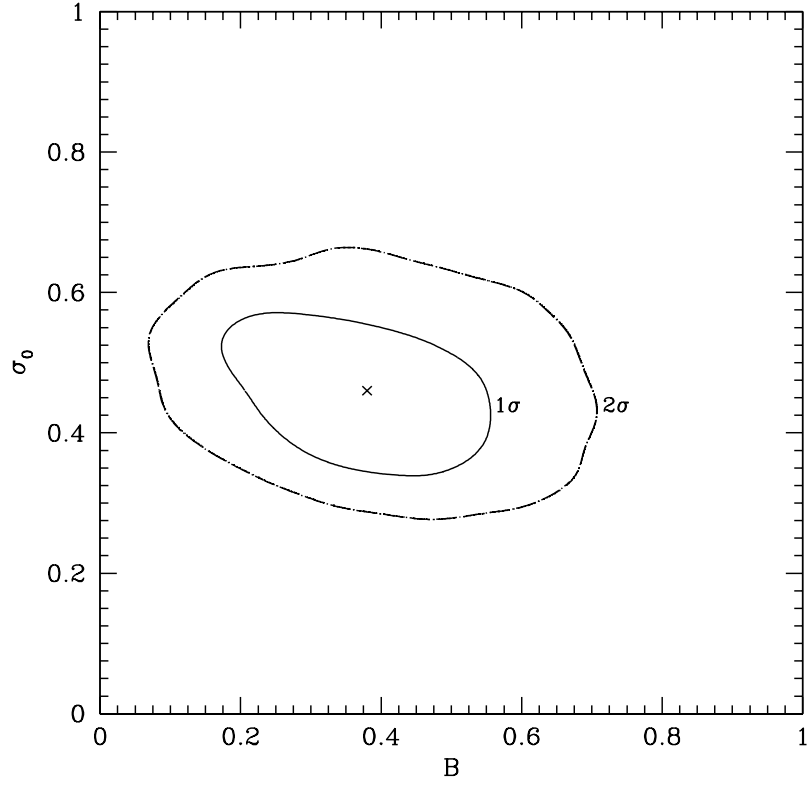}
\vspace{-2cm}
\caption{{Plotted are the 68 and 95 per cent C.L. contours for the slope $B$ and intrinsic scatter $\sigma_0$ which characterize the posterior PDF ${\rm P}_{r} (A,B,\sigma|\{L_{\rm X},M_{\rm BH}\})$ (defined in \S\ \ref{ssec:likeli}, Eq. 1), marginalized over the normalization A. The inferred power law relation between the average X-ray luminosity and black hole mass is of the form: $\log (\langle {L}_{X}\rangle/10^{38}{\rm erg\
s^{-1}}) = A + B \ \log ({\langle {M}_{\rm BH}\rangle}/{10^8 M_{\odot}})$, with
$A=1.0\pm0.1$, $B=0.38^{+0.13}_{-0.12}$, and with an intrinsic scatter of
$0.46^{+0.08}_{-0.06}$ dex. 
\label{fig:pdf}}}}
\end{figure}

\begin{figure}
\center{\includegraphics[angle=0,scale=.6]{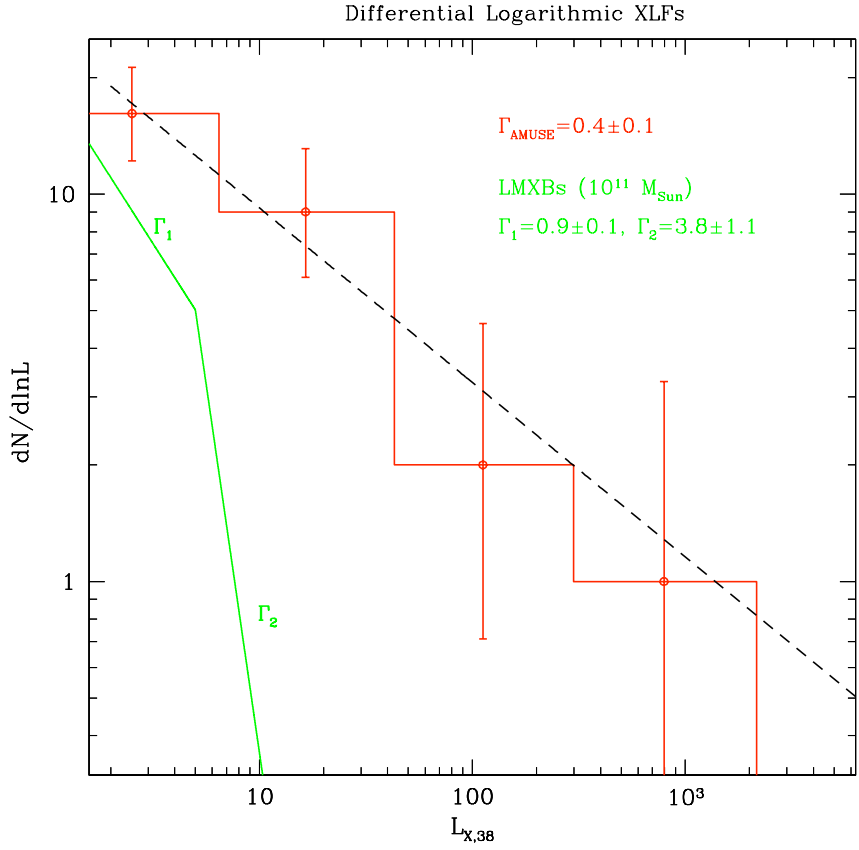}
\vspace{-2cm}
\caption{{\it Left}: The differential logarithmic X-ray luminosity function (XLF) of the accreting SMBHs in AMUSE-Virgo, here shown in red, is compared to the analytic expression for the field LMXB luminosity function given by Gilfanov (2004), in green, with a normalization corresponding to a stellar mass of $10^{11}$ \msun.  The AMUSE-Virgo XLF is well approximated by a power-law with slope $\Gamma=0.4\pm0.1$, much shallower than the slope of the field LMXB XLF at the high luminosity end; this confirms the different physical nature of detected X-ray cores.
\label{fig:xlf}}}
\end{figure}

With the goal of providing an unbiased census of local SMBH activity in early type galaxies, the AMUSE-Virgo survey targets an homogeneous sample of 100 nearby galaxies with the {Chandra X-ray Observatory} and the {\it Spitzer Space Telescope} down to a limiting threshold of \ledd\ for a 3 \msun\ object (3-$\sigma$). 
The targeted sample is that of the \hst\ Virgo Cluster Survey (VCS; C\^ot\'e \etal 2004). For each galaxy, the high resolution $g$ and $z$ band images enable us to resolve, when present, the nuclear star cluster, infer its enclosed mass, and thus estimate the chance contamination from a low-mass X-ray binary (LMXB) as bright/brighter than the detected X-ray core.
The main results and implications of this work can be summarized as follows:

\begin{enumerate}

\item 
Out of 100 objects, 32 show a nuclear X-ray source. After carefully accounting for contamination from LMXBs, making use of their X-ray luminosity function in external galaxies, we are able to conclude that between $24-34\%$  of the galaxies in our sample host a X-ray active SMBH, down to our limiting X-ray luminosity (at the 95$\%$ C.L.). This sets a firm lower limit to the black hole occupation fraction in nearby bulges within a cluster environment.  
\item 
As already reported in Paper I. of this series, we confirm that the fraction $f_X$ of active nuclear SMBHs is an increasing function of the host stellar mass \mstar, with $0.01<f_X<0.14$ for log(\mstar/\msun)$<{9.5}$, $0.53<f_X<0.87$ for log(\mstar/\msun)$>{10.5}$, and $0.16<f_X<0.43$ for intermediate masses (Figure~\ref{fig:activefrac}). While this trend of (fractionally) increasing nuclear SMBH activity with increasing galaxy mass is well known, here we are able to show that it simply results from dealing with a luminosity-limited survey instead of a `Eddington-limited' one. 
\item
Only 6/100 objects host both a nuclear star cluster and a nuclear X-ray source; of those 6, 2 are likely to be heavily contaminated by LMXB X-ray emission. After accounting for this effect, the fraction of such hybrid nuclei is constrained between $0.3-7\%$ for host stellar masses below $10^{11}$  \msun, and to be lower than 32$\%$ above it (95$\%$ C.L.).
\item
The differential logarithmic X-ray luminosity function of active SMBHs in our sample scales with the 
X-ray luminosity as \lx$^{-0.4\pm0.1}$ between a few $10^{38}$ and $10^{42}$ erg \se.
The fitted slope is much shallower than for LMXBs, confirming the different nature of the nuclear X-ray sources' population. 
\item 
In terms of Eddington-scaled luminosities, the inferred ratios range between a few $10^{-9}$ and a few $10^{-6}$, much broader than what reported for high redshift AGN (e.g. Kollmeier \etal 2006). We speculate that it is likely to extend to {\it bona fide} AGN, similarly to what is observed in the hard X-ray state of BH X-ray binaries. 
\item
We use a Bayesian statistical analysis to assess the dependence of
the accretion-powered X-ray luminosity on the black hole mass, taking
into account selection effects. We find that the average \lx/\ledd\
depends on black hole mass as : $\log ({L}_{X}/10^{38}{\rm erg\
s^{-1}}) = A + B \ \log ({{M}_{\rm BH}}/{10^8 M_{\odot}})$, with
$A=1.0\pm0.1$, $B=0.38^{+0.13}_{-0.12}$, and with an intrinsic scatter of
$0.46^{+0.08}_{-0.06}$ dex. In turn, \lx/\ledd $\propto$ \bhm$^{-0.62\pm0.12}$,
arguing for a `down-sizing' behavior of local black hole accretion: low mass black holes shine closer to their Eddington limit than high mass ones.
\item
Stacking the 64 new \cxo\ observations with no visible accreting SMBH results in a statistically significant detection of a nuclear X-ray source with average X-ray luminosity $\langle$\lx/\ledd$\rangle\simeq3.2\pm1.4 \times 10^{37}$ \es, consistent with emission from local LMXBs. This translates into an upper limit to the average Eddington-scaled X-ray luminosity of about $3\times 10^{-8}$, for an average nuclear black hole mass of $10^{7}$ \msun.
\end{enumerate}

\acknowledgments E.G. is supported through Hubble Postdoctoral
Fellowship grant number HST-HF-01218.01-A from the Space Telescope
Science Institute, which is operated by the Association of
Universities for Research in Astronomy, Incorporated, under NASA
contract NAS5-26555. T.T.  acknowledges support from the NSF through
CAREER award NSF-0642621, by the Sloan Foundation through a Sloan
Research Fellowship, and by the Packard Foundation through a Packard
Fellowship. Support for this work was provided by NASA through Chandra
Award Number 08900784 issued by the Chandra X-ray Observatory Center.
We are grateful to Patrick C\^ot\'e\ and Laura Ferrarese for helpful comments, and for providing us with the list of
masses for the nuclear star clusters. We thank the anonymous referee for the thorough and constructive report.

\clearpage
\clearpage
\clearpage
\clearpage
\clearpage
\clearpage

\LongTables 
\begin{center}
\newcommand\tabspace{\noalign{\vspace*{0.7mm}}}
\begin{deluxetable}{llrcrrrrc} 
\setlength{\tabcolsep}{0.05in} 
\tablewidth{0pt} 
\tablecaption{AMUSE-Virgo II.: nuclear properties.\label{tab:cxo}}
\tablehead{      \colhead{ID}
	   &     \colhead{VCC}
	   &     \colhead{Other}
	   &     \colhead{$d$}
	   &     \colhead{B}
	   &     \colhead{\si}
	   &     \colhead{log $M_{\rm BH}$}
	   &     \colhead{log \lx} 
           &    \colhead{log $M_{\star}$}
\\	 
		\colhead{}          
           &       \colhead{}
 	  &       \colhead{}
           &     \colhead{(Mpc)}
	   &     \colhead{(mag)}
  	   &     \colhead{(km \se)}
           &     \colhead{(log \msun)}
	   &     \colhead{[log(erg s$^{-1}$)]}
	   &\colhead{(log \msun)}
           \\	
	         \colhead{(1)} 
	   &	 \colhead{(2)} 
           &     \colhead{(3)} 
           &     \colhead{(4)}  
	   &     \colhead{(5)}  
	   &     \colhead{(6)}  
	   &     \colhead{(7)} 
	   &     \colhead{(8)} 
	   &     \colhead{(9)} 
\\
} 
\startdata 
    1 & 1226 &M49,N4472    &     17.14 &      8.63 &308 $\pm$ 9 &       9.1 & $<$     38.49 &   12.0 \\
    2 & 1316 &M87,N4486    &     17.22 &      9.05 &355 $\pm$ 8 &       9.4 &     41.20 &     11.8 \\ 
    3 & 1978 &M60,N4649    &     17.30 &      9.33 &347 $\pm$ 9 &       9.4 &     39.05 &     11.7 \\
    4 &  881 &M86,N4406    &     16.83 &      8.77 &245 $\pm$11 &       8.6 &$<$ 38.64 &      11.9 \\
    5 &  798 &M85,N4382    &     17.86 &      9.30 &205 $\pm$ 8 &       8.3 &$<$ 38.43 &     11.6 \\
    6 &  763 &M84,N4374    &     18.45 &      9.35 &297 $\pm$ 7 &       9.1 &     39.73 &      11.7 \\
    7 &  731 &N4365       &     23.33 &      9.99 &261 $\pm$ 7 &       8.8 &     39.00 &      11.7 \\
    8 & 1535 &N4526       &     16.50 &     10.52 &316 $\pm$ 7 &       9.2  &$<$ 38.21 &      11.0 \\
    9 & 1903 &M59,N4621    &     14.93 &     10.02 &233 $\pm$ 7 &       8.5 &     39.11 &      11.3 \\
   10 & 1632 &M89,N4552    &     15.85 &     10.13 &257 $\pm$18 &       8.7 &     39.58 &      11.3 \\
   11 & 1231 &N4473       &     15.28 &     11.19 &189 $\pm$10 &       8.1 &     38.60 &      10.8 \\
   12 & 2095 &N4762       &     16.50 &     11.97 &147 $\pm$10 &       7.6 &     38.71 &      10.6 \\
   13 & 1154 &N4459       &     16.07 &     11.07 &170 $\pm$12 &       7.9 &     39.03 &      10.9\\
   14 & 1062 &N4442       &     15.28 &     11.32 &197 $\pm$15 &       8.2 &     38.47 &      10.7 \\
   15 & 2092 &N4754       &     16.14 &     11.36 &200 $\pm$10 &       8.2 &     38.59 &      10.9 \\
   16 &  369 &N4267       &     15.85 &     12.39 &165 $\pm$11 &       7.8 &     39.26 &      10.4 \\
   17 &  759 &N4371       &     16.98 &     11.55 &129 $\pm$ 6 &       7.3 & $<$ 38.18 &      10.8\\
   18 & 1692 &N4570       &     17.06 &     11.91 &180 $\pm$18 &       8.0 &     38.45 &      10.6 \\
   19 & 1030 &N4435       &     16.75 &     11.75 &174 $\pm$16 &       7.9 &     38.72 &      10.8 \\
   20 & 2000 &N4660       &     15.00 &     11.93 &203 $\pm$ 5 &       8.3 &     38.65 &      10.4\\
   21 &  685 &N4350       &     16.50 &     11.83 &198 $\pm$ 9 &       8.2 &     39.14 &      10.6\\
   22 & 1664 &N4564       &     15.85 &     11.85 &157 $\pm$ 9 &       7.7 &     39.95 &      10.6 \\
   23 &  654 &N4340       &     16.50 &     12.22 &112 $\pm$ 3 &       7.0 & $<$     38.46 &      10.4\\
   24 &  944 &N4417       &     16.00 &     12.06 &125 $\pm$ 4 &       7.2 & $<$     38.53 &      10.4 \\
   25 & 1938 &N4638       &     17.46 &     12.01 &132 $\pm$10 &       7.3 &     38.97 &      10.5 \\
   26 & 1279 &N4478       &     16.98 &     12.21 &147 $\pm$ 6 &       7.6 & $<$     38.73 &      10.5 \\
   27 & 1720 &N4578       &     16.29 &     12.01 &153 $\pm$15 &       7.7 & $<$     38.54 &      10.4     \\
   28 &  355 &N4262       &     15.42 &     12.29 &179 $\pm$22 &       8.0 &     38.77 &      10.3 \\
   29 & 1619 &N4550       &     15.49 &     12.37 & 93 $\pm$ 5 &       6.6 &     38.68 &      10.2\\
   30 & 1883 &N4612       &     16.60 &     12.01 &104 $\pm$11 &       6.8 &     38.35 &      10.4\\
   31 & 1242 &N4474       &     15.56 &     12.38 & 93 $\pm$ 7 &       6.6 & $<$     38.50 &      10.3 \\
   32 &  784 &N4379       &     15.85 &     12.44 &103 $\pm$ 5 &       6.8 &     38.62 &      10.3 \\
   33 & 1537 &N4528       &     15.85 &     12.73 &112 $\pm$ 9 &       7.0 &     38.52 &      10.1 \\
   34 &  778 &N4377       &     17.78 &     12.87 &139 $\pm$17 &       7.5 &     38.56 &      10.2 \\
   35 & 1321 &N4489       &     15.42 &     12.51 & ... &       7.7 & $<$     38.33 &      10.1 \\
   36 &  828 &N4387       &     17.95 &     12.93 & 99 $\pm$ 4 &       6.7 & $<$     38.63 &      10.2\\
   37 & 1250 &N4476       &     17.62 &     12.63 & ... &       7.8 &     38.73 &      10.2 \\
   38 & 1630 &N4551       &     16.14 &     12.74 &102 $\pm$ 5 &       6.8 & $<$     38.29 &      10.2\\
   39 & 1146 &N4458       &     16.37 &     12.95 &110 $\pm$ 8 &       7.0 & $<$     38.33 &      10.0 \\
   40 & 1025 &N4434       &     22.44 &     12.87 &119 $\pm$ 6 &       7.1 &     38.92 &      10.4 \\
   41 & 1303 &N4483       &     16.75 &     12.94 &106 $\pm$10 &       6.9 & $<$     38.15 &      10.1\\
   42 & 1913 &N4623       &     17.38 &     13.16 & 89 $\pm$10 &       6.5 & $<$     38.46 &      10.1 \\
   43 & 1327 &N4486A      &     18.28 &     13.25 & ... &       7.6 &     38.68 &      10.1 \\
   44 & 1125 &N4452       &     16.50 &     13.07 &114 $\pm$ 7 &       7.0 & $<$     38.46 &       9.9 \\
   45 & 1475 &N4515       &     16.60 &     13.28 & 91 $\pm$10 &       6.6 & $<$     38.34 &       9.9\\
   46 & 1178 &N4464       &     15.85 &     13.32 &121 $\pm$25 &       7.2 &     38.66 &       9.9\\
   47 & 1283 &N4479       &     17.38 &     13.25 & 82 $\pm$ 7 &       6.3 &     38.54 &      10.1 \\
   48 & 1261 &N4482       &     18.11 &     13.52 & 45 $\pm$ 0 &       5.1 & $<$     38.42 &       9.8 \\
   49 &  698 &N4352       &     18.71 &     13.28 & 85 $\pm$ 8 &       6.4 & $<$     38.44 &      10.0 \\
   50 & 1422 &I3468       &     15.35 &     13.82 & ... &       7.2 & $<$     38.08 &       9.6 \\
   51 & 2048 &I3773       &     16.50 &     14.04 & 79 $\pm$ 5 &       6.3 & $<$     38.12 &       9.6\\
   52 & 1871 &I3653       &     15.49 &     14.36 & ... &       6.9 & $<$     38.08 &       9.5 \\
   53 &    9 &I3019         &     17.14 &     14.01 & ... &       7.2 & $<$     38.15 &       9.7\\
   54 &  575 &N4318       &     22.08 &     11.69 & 95 $\pm$ 4 &       6.6 & $<$     38.39 &      10.8 \\
   55 & 1910 &I809        &     16.07 &     14.34 & ... &       7.0 & $<$     38.30 &       9.5 \\
   56 & 1049 &U7580       &     16.00 &     14.93 & ... &       6.7 & $<$     38.08 &       9.0\\
   57 &  856 &I3328       &     16.83 &     14.50 & 34 $\pm$ 1 &       4.5 & $<$     38.16 &       9.5 \\
   58 &  140 &I3065       &     16.37 &     14.28 & ... &       7.0 & $<$     38.13 &       9.4 \\
   59 & 1355 &I3442       &     16.90 &     14.65 & ... &       6.9 &     38.58 &       9.4\\
   60 & 1087 &I3381       &     16.67 &     14.11 & 38 $\pm$ 1 &       4.8 & $<$     38.15 &       9.6 \\
   61 & 1297 &N4486B      &     16.29 &     14.15 &166 $\pm$ 8 &       7.8 &     38.42 &       9.7 \\
   62 & 1861 &I3652       &     16.14 &     14.45 & ... &       6.9 & $<$     38.12 &       9.5 \\
   63 &  543 &U7436       &     15.70 &     14.34 & 27 $\pm$ 0 &       4.1 & $<$     38.27 &       9.4 \\
   64 & 1431 &I3470       &     16.14 &     14.44 & ... &       6.9 & $<$     38.66 &       9.5 \\
   65 & 1528 &I3501       &     16.29 &     14.57 & ... &       6.9 & $<$     38.13 &       9.3 \\
   66 & 1695 &I3586       &     16.52 &     14.40 & ... &       7.0 & $<$     38.14 &       9.5\\
   67 & 1833 &...         &     16.22 &     14.66 & ... &       6.9 & $<$     38.11 &       9.3\\
   68 &  437 &U7399A      &     17.14 &     14.05 & ... &       7.2 & $<$     38.17 &       9.6 \\
   69 & 2019 &I3735       &     17.06 &     14.68 & ... &       6.9 & $<$     38.17 &       9.4 \\
   70 &   33 &I3032       &     15.07 &     15.23 & ... &       6.6 & $<$     38.25 &       8.9 \\
   71 &  200 &...         &     18.20 &     15.01 & ... &       6.8 & $<$     38.42 &       9.2\\
   72 &  571 &...         &     23.77 &     15.02 & ... &       7.1 & $<$     38.46 &       9.4 \\
   73 &   21 &I3025       &     16.50 &     15.04 & ... &       6.7 & $<$     38.11 &       9.0 \\
   74 & 1488 &I3487       &     16.50 &     15.03 & 28 $\pm$ 0 &       4.1 & $<$     38.14 &       9.0\\
   75 & 1779 &I3612       &     16.50 &     14.97 & ... &       6.7 & $<$     38.14 &       9.0 \\
   76 & 1895 &U7854       &     15.85 &     15.15 & ... &       6.6 & $<$     38.10 &       9.0 \\
   77 & 1499 &I3492       &     16.50 &     15.15 & ... &       6.7 &     38.42 &       8.8\\
   78 & 1545 &I3509       &     16.83 &     14.95 & ... &       6.8 & $<$     38.16 &       9.2 \\
   79 & 1192 &N4467       &     16.50 &     14.74 & 83 $\pm$10 &       6.4 & $<$     38.68 &       9.5 \\
   80 & 1857 &I3647       &     16.50 &     15.06 & ... &       6.7 & $<$     38.14 &       9.0\\
   81 & 1075 &I3383       &     16.14 &     15.20 & ... &       6.6 & $<$     38.12 &       9.2 \\
   82 & 1948 &...         &     16.50 &     15.76 & ... &       6.4 & $<$     38.14 &       8.8 \\
   83 & 1627 &...         &     15.63 &     15.31 & ... &       6.6 & $<$     38.34 &       9.1 \\
   84 & 1440 &I798        &     16.00 &     14.90 & ... &       6.7 & $<$     38.27 &       9.2 \\
   85 &  230 &I3101       &     17.78 &     15.68 & ... &       6.5 & $<$     38.62 &       8.9\\
   86 & 2050 &I3779       &     15.78 &     15.37 & ... &       6.5 & $<$     38.10 &       9.0 \\
   87 & 1993 &...         &     16.52 &     15.79 & ... &       6.4 & $<$     38.14 &       9.0\\
   88 &  751 &I3292       &     15.78 &     14.86 & ... &       6.7 & $<$     38.29 &       9.4\\
   89 & 1828 &I3635       &     16.83 &     15.29 & ... &       6.6 & $<$     38.16 &       9.1 \\
   90 &  538 &N4309A      &     22.91 &     16.17 & ... &       6.5 & $<$     38.41 &       8.9 \\
   91 & 1407 &I3461       &     16.75 &     15.25 & ... &       6.6 & $<$     38.35 &       9.1 \\
   92 & 1886 &...         &     16.50 &     15.50 & ... &       6.5 & $<$     38.14 &       8.8 \\
   93 & 1199 &...         &     16.50 &     16.00 & ... &       6.3 & $<$     38.14 &       9.0 \\
   94 & 1743 &I3602       &     17.62 &     15.73 & ... &       6.5 & $<$     38.20 &       8.9 \\
   95 & 1539 &...         &     16.90 &     15.36 & ... &       6.6 & $<$     38.16 &       8.9 \\
   96 & 1185 &...         &     16.90 &     15.50 & ... &       6.5 & $<$     38.36 &       9.1 \\
   97 & 1826 &I3633       &     16.22 &     15.78 & ... &       6.4 & $<$     38.12 &       8.8\\
   98 & 1512 &...         &     18.37 &     13.92 & ... &       7.3 & $<$     38.23 &       9.2 \\
   99 & 1489 &I3490       &     16.50 &     16.10 & ... &       6.3 & $<$     38.14 &       8.7\\
  100 & 1661 &...         &     15.85 &     14.80 & ... &       6.8 & $<$     38.12 &       9.0 \\
\\
\enddata 
\tablecomments{Col.: (1) ACSVCS target number (2) VCC source name; (3)
Alternate name, from NCG or catalogs; (4)
Distance (from surface brightness fluctuations method; Mei \etal
2007; see also Blakeslee et al. 2009). The average distance to the Virgo cluster -- of 16.5 Mpc -- is
employed in case of no available distance modulus; (5)
Extinction-corrected B magnitude, estimated as described in
paper I; E(B-V) values are from Ferrarese \etal (2006b);
(6) Stellar velocity dispersion, from ENEARc (Bernardi \etal 2002), unless
otherwise indicated (for details, see caption of Table 2 in Paper I.); (7) Black hole mass, estimated according to the `fiducial distribution' described in \S\ 4. of Paper I.;
(8) Nuclear luminosity between
0.3-10 keV, corrected for absorption; literature references are given in
brackets; (9) Stellar mass of the host galaxy, in \msun, derived from $g_0$ and
$z_0$ band AB model magnitudes following Bell \etal (2003), as
described in Paper I.  }
\end{deluxetable} 
\end{center}

\clearpage
\begin{center}
\begin{deluxetable}{lrrcrrcrr} 
\setlength{\tabcolsep}{0.05in} 
\tablecaption{Galaxies hosting both a nuclear X-ray source and a massive nuclear star cluster.\label{tab:nsc}}
\tablehead{   \colhead{ID}
	   &     \colhead{VCC}
	   &     \colhead{$g$}
	   &     \colhead{$z$}
	   &     \colhead{$r_h$}
	   &     \colhead{log $M_{\rm s}$}
	   &     \colhead{N($>$\lx)}
	   &     \colhead{$1-P_X$} &
\\	 
		\colhead{}          
           &       \colhead{}
 	  &       \colhead{(mag)}
           &     \colhead{(mag)}
	   &     \colhead{(\arcsec)}
  	   &     \colhead{(\msun)}
	   &	\colhead{}
           &     \colhead{($\%$)}
           \\	
	         \colhead{(1)} 
	   &	 \colhead{(2)} 
	   &     \colhead{(3)}
             &     \colhead{(4)} 
	   &     \colhead{(5)}  
	   &     \colhead{(6)}  
	   &     \colhead{(7)}
	   & \colhead{(8)}
} 
\startdata 

    29 & 1619 & 17.15 &     15.61 &  0.324 & 8.1     & 0.42 & 34.29\\
    30 & 1883 &  18.75   &     17.63  &    0.024  & 7.3   & 8.33 & 99.97   \\
    32 & 784 &  18.36  & 16.71  &    0.161 &  7.7 &  0.72 & 51.32\\
    37 & 1250 &  19.75  & 18.22    &    0.026  &  7.2   & 8.59 & 99.98 \\
    47 &  1283 &   20.67 &  19.10   & 0.053   &   6.8     & 0.59 & 44.57 \\
    59 &  1355 &  21.11  &  20.10   &  0.043    &   6.3    & 0.11 & 10.42\\
\\
\enddata 
\tablecomments{Columns: (1) \& (2) see Table 1; (3) \& (4) $g$- and $z$- band magnitudes, from Ferrarese \etal (2006b); (5) \& (6) half-light radius (from Ferrarese \etal 2006) and massive star cluster mass estimated following the prescription and mass-to-light ratio described in Ferrarese \etal (2006a); (7) Number of expected LMXBs with X-ray luminosity equal or greater than the measured X-ray core; (8) $P_X$ is the chance probability that the \cxo\ PSF is contaminated by a LMXB of luminosity equal/higher than the measured one.\\}
\end{deluxetable} 
\end{center}
\clearpage
\begin{center}
\begin{deluxetable}{cccll} 
\setlength{\tabcolsep}{0.05in} 
\tablecaption{X-ray active fraction \& completeness.\label{tab:completeness}}
\tablehead{   \colhead{\lx/\ledd}
	   &     \colhead{$\#$ objects}
	   &     \colhead{10$^{8.5-10.5}$\msun}
	   &     \colhead{10$^{10.5-12.0}$\msun}
\\	 
	         \colhead{(1)} 
	   &	 \colhead{(2)} 
           &     \colhead{(3)} 
           &     \colhead{(4)}  
} 
\startdata 
-9 & 4 & 0--100$\%$ (0/0) & 25.1--98.8$\%$ (3/4)\\
-8 & 11 & $>4.5\%$ (1/1) & 60.2--99.5$\%$ (9/10)\\
-7 & 13 & 19.3--80.6$\%$ (4/8) &  $>48.3\%$ (5/5)\\
-6 & 3 & $<88.2\%$ (0.41$^a$/2)& $>4.5\%$ (1/1)\\
 
\enddata 
\tablecomments{Columns: (1) Completeness limit in Eddington-scaled X-ray luminosity. (2) Total number of objects within the given completeness limit. (3) X-ray active fraction for host stellar masses ranging between 10$^{8.5-10.5}$ \msun. The number of active vs. total is given in parenthesis. (4) X-ray active fraction for host stellar masses ranging between 10$^{10.5-12.0}$ \msun. The number of active vs. total is given in parenthesis. $a$) Refers to the `weighted' SMBH distribution (see \S~\ref{sec:lmxb}).}
\end{deluxetable} 
\end{center}


\begin{thebibliography}

\bibitem[]{}
Alexander, D. M., \etal 2008, \aj, 135, 1968

\bibitem[{{Allen} {et~al.}(2000){Allen}, {Di Matteo}, \& {Fabian}}]{ADF00}
{Allen}, S.~W., {Di Matteo}, T., \& {Fabian}, A.~C. 2000, \mnras, 311, 493

\bibitem[{{Bell} {et~al.}(2003){Bell}, {McIntosh}, {Katz}, \&
  {Weinberg}}]{Bel++03}
{Bell}, E.~F., {McIntosh}, D.~H., {Katz}, N., \& {Weinberg}, M.~D. 2003, \apjs,
  149, 289
  
\bibitem[]{}
Bentz, M., \etal 2009, \apj, 705, 199

\bibitem[{{Bernardi} {et~al.}(2002){Bernardi}, {Alonso}, {da Costa}, {Willmer},
  {Wegner}, {Pellegrini}, {Rit{\'e}}, \& {Maia}}]{Ber++02}
{Bernardi}, M., {Alonso}, M.~V., {da Costa}, L.~N., {Willmer}, C.~N.~A.,
  {Wegner}, G., {Pellegrini}, P.~S., {Rit{\'e}}, C., \& {Maia}, M.~A.~G. 2002,
  \aj, 123, 2990
  
\bibitem[]{}
Blakeslee, J. P., et al. 2009, \apj, 694, 556
  
\bibitem[]{}
Bundy, K., et al. 2008, ApJ, 681, 931
  
\bibitem[]{}
Canizares, C. R., Fabbiano, G., \& Trinchieri, G. 1997, \apj, 312, 503
  
\bibitem[]{}
Choi, W. W., Woo, J.-H., \& Park, C. 2009, \apj, 699, 1679

\bibitem[{{Ciotti} \& {Ostriker}(2007)}]{C+O07}
{Ciotti}, L., \& {Ostriker}, J.~P. 2007, \apj, 665, 1038

\bibitem[]{}
Constantin, A., Hoyle, F., \& Vogeley, M. S. 2008, ApJ, 673, 715

\bibitem[]{}
Constantin, A., \& Vogeley, M. S. 2006, \apj, 650, 727

\bibitem[]{}
{C{\^o}t{\'e}}, P., \etal 2007, \apj, 671, 1456,

\bibitem[{{C{\^o}t{\'e}} {et~al.}(2006){C{\^o}t{\'e}}, {Piatek}, {Ferrarese},
  {Jord{\'a}n}, {Merritt}, {Peng}, {Ha{\c s}egan}, {Blakeslee}, {Mei}, {West},  {Milosavljevi{\'c}}, \& {Tonry}}]{C++06}
{C{\^o}t{\'e}}, P., {Piatek}, S., {Ferrarese}, L., {Jord{\'a}n}, A., {Merritt},
  D., {Peng}, E.~W., {Ha{\c s}egan}, M., {Blakeslee}, J.~P., {Mei}, S., {West},
  M.~J., {Milosavljevi{\'c}}, M., \& {Tonry}, J.~L. 2006, \apjs, 165, 57

\bibitem[{{C{\^o}t{\'e}} {et~al.}(2004){C{\^o}t{\'e}}, {Blakeslee},
  {Ferrarese}, {Jord{\'a}n}, {Mei}, {Merritt}, {Milosavljevi{\'c}}, {Peng},
  {Tonry}, \& {West}}]{C++04}
{C{\^o}t{\'e}}, P., {Blakeslee}, J.~P., {Ferrarese}, L., {Jord{\'a}n}, A.,
  {Mei}, S., {Merritt}, D., {Milosavljevi{\'c}}, M., {Peng}, E.~W., {Tonry},
  J.~L., \& {West}, M.~J. 2004, \apjs, 153, 223

\bibitem[]{}
Cowie, L. L., Songaila, A., Hu, E. M., \& Cohen, J. G. 1996, \aj, 112, 839

\bibitem[{{Decarli} {et~al.}(2007){Decarli}, {Gavazzi}, {Arosio}, {Cortese},
  {Boselli}, {Bonfanti}, \& {Colpi}}]{Dec++07}
{Decarli}, R., {Gavazzi}, G., {Arosio}, I., {Cortese}, L., {Boselli}, A.,
  {Bonfanti}, C., \& {Colpi}, M. 2007, \mnras, 381, 136

\bibitem[{{Dickey} \& {Lockman}(1990)}]{D+L90}
{Dickey}, J.~M., \& {Lockman}, F.~J. 1990, \araa, 28, 215

\bibitem[{{Di Matteo} {et~al.}(2003){Di Matteo}, {Allen}, {Fabian}, {Wilson},
  \& {Young}}]{DiM++03}
{Di Matteo}, T., {Allen}, S.~W., {Fabian}, A.~C., {Wilson}, A.~S., \& {Young},
  A.~J. 2003, \apj, 582, 133

\bibitem[]{}
Di Matteo, T., Carilli, C. L., \& Fabian, A. C. 2001, \mnras, 547, 731

\bibitem[{{Di Matteo} {et~al.}(2000){Di Matteo}, {Quataert}, {Allen},
  {Narayan}, \& {Fabian}}]{DiM++00}
{Di Matteo}, T., {Quataert}, E., {Allen}, S.~W., {Narayan}, R., \& {Fabian},
  A.~C. 2000, \mnras, 311, 507

\bibitem[]{}
Fabbiano, G., Baldi, A., Pellegrini, S., Siemiginowska, A., Elvis, M., Zezas, A., \& McDowell, J. 2004, \apj, 616, 730

\bibitem[]{}
Fabbiano, G., Elvis, M., Markoff, S., Siemiginowska, A., Pellegrini, S., Zezas, A., Nicastro, F., Trinchieri, G., \& McDowell, J. 2003, \apj, 588, 175

\bibitem[]{}
Fabbiano, G. 1993, Advances in Space Research, 12, 161

\bibitem[{{Fabbiano} \& {Juda}(1997)}]{F+J97}
{Fabbiano}, G., \& {Juda}, J.~Z. 1997, \apj, 476, 666

\bibitem[{{Ferrarese} {et~al.}(2006){Ferrarese}, {C{\^o}t{\'e}}, {Dalla
  Bont{\`a}}, {Peng}, {Merritt}, {Jord{\'a}n}, {Blakeslee}, {Ha{\c s}egan},
  {Mei}, {Piatek}, {Tonry}, \& {West}}]{Fer++06a}
{Ferrarese}, L., {C{\^o}t{\'e}}, P., {Dalla Bont{\`a}}, E., {Peng}, E.~W.,
  {Merritt}, D., {Jord{\'a}n}, A., {Blakeslee}, J.~P., {Ha{\c s}egan}, M.,
  {Mei}, S., {Piatek}, S., {Tonry}, J.~L., \& {West}, M.~J. 2006a, \apjl, 644,
  L21

\bibitem[{{Ferrarese} {et~al.}(2006){Ferrarese}, {C{\^o}t{\'e}}, {Jord{\'a}n},
  {Peng}, {Blakeslee}, {Piatek}, {Mei}, {Merritt}, {Milosavljevi{\'c}},
  {Tonry}, \& {West}}]{Fer++06b}
{Ferrarese}, L., {C{\^o}t{\'e}}, P., {Jord{\'a}n}, A., {Peng}, E.~W.,
  {Blakeslee}, J.~P., {Piatek}, S., {Mei}, S., {Merritt}, D.,
  {Milosavljevi{\'c}}, M., {Tonry}, J.~L., \& {West}, M.~J. 2006b, \apjs, 164,
  334

\bibitem[{{Ferrarese} \& {Ford}(2005)}]{F+F05}
{Ferrarese}, L., \& {Ford}, H. 2005, Space Science Reviews, 116, 523

\bibitem[{{Ferrarese} \& {Merritt}(2000)}]{F+M00}
{Ferrarese}, L., \& {Merritt}, D. 2000, \apjl, 539, L9

\bibitem[]{}
Filippenko, A. V., \& Ho, L. C., 2003, \apj, 588, L13

\bibitem[]{}
Gallo, E., Treu, T., Jacob, J., Woo, J.-H., Marshall, P. J., \& Antonucci, R. 2008, \apj, 680, 154 (Paper I)

\bibitem[{{Garmire} {et~al.}(2000){Garmire}, {Feigelson}, {Broos},
  {Hillenbrand}, {Pravdo}, {Townsley}, \& {Tsuboi}}]{Gar++00}
{Garmire}, G., {Feigelson}, E.~D., {Broos}, P., {Hillenbrand}, L.~A., {Pravdo},
  S.~H., {Townsley}, L., \& {Tsuboi}, Y. 2000, \aj, 120, 1426

\bibitem[{{Gebhardt} {et~al.}(2000){Gebhardt}, {Kormendy}, {Ho}, {Bender},
  {Bower}, {Dressler}, {Faber}, {Filippenko}, {Green}, {Grillmair}, {Lauer},
  {Magorrian}, {Pinkney}, {Richstone}, \& {Tremaine}}]{Geb++00}
{Gebhardt}, K., {Kormendy}, J., {Ho}, L.~C., {Bender}, R., {Bower}, G.,
  {Dressler}, A., {Faber}, S.~M., {Filippenko}, A.~V., {Green}, R.,
  {Grillmair}, C., {Lauer}, T.~R., {Magorrian}, J., {Pinkney}, J., {Richstone},  D., \& {Tremaine}, S. 2000, \apjl, 543, L5

\bibitem[{{Gehrels}(1986)}]{Geh86}
{Gehrels}, N. 1986, \apj, 303, 336

\bibitem[]{}
Georgakakis, A., et al. 2008, \mnras, 385, 204

\bibitem[]{}	
Ghosh, H. Mathur, S., Fiore, F., \& Ferrarese, L. 2008, \apj, 687, 216

\bibitem[{{Gilfanov}(2004)}]{Gil04}
{Gilfanov}, M. 2004, \mnras, 349, 146

\bibitem[]{}	
Gonz\'alez Delgado, R. M., P\'erez, E., Cid Fernandes, R., \& Schmitt, H., 2008, \apj, 135, 747

\bibitem[{{Graham}(2009)}]{Gra09}
{Graham}, A.~W., \& Spitler, L. R. 2009, \mnras, 397, 2148

\bibitem[{{Graham}(2007)}]{Gra07}
Graham, A. W., \& Driver, S. P. 2007, \mnras, 380, L15

\bibitem[{{Greene} \& {Ho}(2009)}]{G+H09}
{Greene}, J.~E., \& {Ho}, L.~C. 2009 \apj, 704, 1743

\bibitem[{{Greene} \& {Ho}(2007)}]{G+H07a}
{Greene}, J.~E., \& {Ho}, L.~C. 2007, \apj, 667, 131

\bibitem[]{}
G\"ultekin, K., \etal 2009, \apj, 698, 198

\bibitem[{{Heckman} {et~al.}(2004){Heckman}, {Kauffmann}, {Brinchmann},
  {Charlot}, {Tremonti}, \& {White}}]{Hec++04}
{Heckman}, T.~M., {Kauffmann}, G., {Brinchmann}, J., {Charlot}, S., {Tremonti},
  C., \& {White}, S.~D.~M. 2004, \apj, 613, 109

\bibitem[]{}
Ho, L.~C. 2008, \araa, 46, 475

\bibitem[{{Ho} {et~al.}(2001){Ho}, {Feigelson}, {Townsley}, {Sambruna},
  {Garmire}, {Brandt}, {Filippenko}, {Griffiths}, {Ptak}, \&
  {Sargent}}]{Ho++01}
{Ho}, L.~C., {Feigelson}, E.~D., {Townsley}, L.~K., {Sambruna}, R.~M.,
  {Garmire}, G.~P., {Brandt}, W.~N., {Filippenko}, A.~V., {Griffiths}, R.~E.,
  {Ptak}, A.~F., \& {Sargent}, W.~L.~W. 2001, \apjl, 549, L51

\bibitem[]{}
Ho, L.~C., Filippenko, A. V., \& Sargent, W. L. W. 1997, \aj, 487, 568

\bibitem[]{}
Humphrey, P. J., \& Buote, D. A. 2008, \apj, 689, 983

\bibitem[]{}
Kauffmann, G., Heckman, T. M., \& Best, P. N. 2008, \mnras, 353,  713

\bibitem[]{}
Kauffmann, G., \etal 2004, \mnras, 353,  713

\bibitem[]{}
Kauffmann, G., \etal 2003, \mnras, 346, 1055

\bibitem[]{}
Kewley, L. J., Groves, B., Kauffmann, G., \& Heckman, T. 2006, \mnras, 372, 961

\bibitem[]{}
Kim, D.-W., \& Fabbiano, G. 2004, \apj, 611, 846

\bibitem[{{Kollmeier} {et~al.}(2006){Kollmeier}, {Onken}, {Kochanek}, {Gould},
  {Weinberg}, {Dietrich}, {Cool}, {Dey}, {Eisenstein}, {Jannuzi}, {Le Floc'h},
  \& {Stern}}]{Kol++06}
{Kollmeier}, J.~A., {Onken}, C.~A., {Kochanek}, C.~S., {Gould}, A., {Weinberg},
  D.~H., {Dietrich}, M., {Cool}, R., {Dey}, A., {Eisenstein}, D.~J., {Jannuzi},
  B.~T., {Le Floc'h}, E., \& {Stern}, D. 2006, \apj, 648, 128

\bibitem[]{}
Kormendy, J., Fisher, D. B., Cornell, M. E., \& Bender, R. 2009, \apjs, 182, 216

\bibitem[{{Kormendy} {et~al.}(1997){Kormendy}, {Bender}, {Magorrian},
  {Tremaine}, {Gebhardt}, {Richstone}, {Dressler}, {Faber}, {Grillmair}, \&
  {Lauer}}]{Kor++97}
{Kormendy}, J., {Bender}, R., {Magorrian}, J., {Tremaine}, S., {Gebhardt}, K.,
  {Richstone}, D., {Dressler}, A., {Faber}, S.~M., {Grillmair}, C., \& {Lauer},
  T.~R. 1997, \apjl, 482, L139

\bibitem[{{Kormendy} \& {Richstone}(1995)}]{K+R95}
{Kormendy}, J., \& {Richstone}, D. 1995, \araa, 33, 581

\bibitem[{{Lauer} {et~al.}(2007){Lauer}, {Gebhardt}, {Faber}, {Richstone},
  {Tremaine}, {Kormendy}, {Aller}, {Bender}, {Dressler}, {Filippenko}, {Green},
  \& {Ho}}]{Lau++07}
{Lauer}, T.~R., {Gebhardt}, K., {Faber}, S.~M., {Richstone}, D., {Tremaine},
  S., {Kormendy}, J., {Aller}, M.~C., {Bender}, R., {Dressler}, A.,
  {Filippenko}, A.~V., {Green}, R., \& {Ho}, L.~C. 2007, \apj, 664, 226

\bibitem[{{Lauer} {et~al.}(1996){Lauer}, {Tremaine}, {Ajhar}, {Bender},
  {Dressler}, {Faber}, {Gebhardt}, {Grillmair}, {Kormendy}, \&
  {Richstone}}]{Lau++96}
{Lauer}, T.~R., {Tremaine}, S., {Ajhar}, E.~A., {Bender}, R., {Dressler}, A.,
  {Faber}, S.~M., {Gebhardt}, K., {Grillmair}, C.~J., {Kormendy}, J., \&
  {Richstone}, D. 1996, \apjl, 471, L79

\bibitem[{{Loewenstein} {et~al.}(2001){Loewenstein}, {Mushotzky}, {Angelini},
  {Arnaud}, \& {Quataert}}]{Loe++01}
{Loewenstein}, M., {Mushotzky}, R.~F., {Angelini}, L., {Arnaud}, K.~A., \&
  {Quataert}, E. 2001, \apjl, 555, L21

\bibitem[]{}
Magorrian, J., \etal 1997, \aj, 115, 2285

\bibitem[{{Marconi} {et~al.}(2004){Marconi}, {Risaliti}, {Gilli}, {Hunt},
  {Maiolino}, \& {Salvati}}]{Mar++04}
{Marconi}, A., {Risaliti}, G., {Gilli}, R., {Hunt}, L.~K., {Maiolino}, R., \&
  {Salvati}, M. 2004, \mnras, 222, 49

\bibitem[{{Marconi} \& {Hunt}(2003)}]{M+H03}
{Marconi}, A., \& {Hunt}, L.~K. 2003, \apjl, 589, L21

\bibitem[{{McLure} \& {Dunlop}(2002)}]{M+D02}
{McLure}, R.~J., \& {Dunlop}, J.~S. 2002, \mnras, 331, 795

\bibitem[{{Mei} {et~al.}(2007){Mei}, {Blakeslee}, {C{\^o}t{\'e}}, {Tonry},
  {West}, {Ferrarese}, {Jord{\'a}n}, {Peng}, {Anthony}, \& {Merritt}}]{Mei++07}
{Mei}, S., {Blakeslee}, J.~P., {C{\^o}t{\'e}}, P., {Tonry}, J.~L., {West},
  M.~J., {Ferrarese}, L., {Jord{\'a}n}, A., {Peng}, E.~W., {Anthony}, A., \&
  {Merritt}, D. 2007, \apj, 655, 144

\bibitem[]{}
Merloni, A., \& Heinz, S. 2008, \mnras, 388, 1011

\bibitem[]{}
Miller, C. J., Nichol, R. C., G\'omez, P. L., Hopkins, A. M., \& Bernardi, M. 2003, \apj, 597, 142

\bibitem[]{}
Nandra, K., et al. 2007, \apj, 660, L11

\bibitem[]{}
Ott, J., Walter, F., \& Brinks, E. 2005, \mnras, 358, 1453

\bibitem[]{}
Page, M. J., Stevens, J. A., Ivison, R. J., \& Carrera, F. J. 2004, \apj, 611, L85

\bibitem[]{}
Pellegrini, S., Ciotti, L., \& Ostriker, J. P. 2007, \apj, 4, 340

\bibitem[{{Pellegrini}(2005)}]{Pel05}
{Pellegrini}, S. 2005, \apj, 624, 155

\bibitem[]{}
Rich, R. M. 2005, \apj, 619, L107

\bibitem[]{}
Roberts, T. P., \& Warwick, R. S. 2000, \mnras, 315, 98

\bibitem[]{}
Salim, S., et al. 2007, \apjs, 173, 267

\bibitem[{{Santra} {et~al.}(2007){Santra}, {Sanders}, \& {Fabian}}]{SSF07}
{Santra}, S., {Sanders}, J.~S., \& {Fabian}, A.~C. 2007, \mnras, 382, 895

\bibitem[]{}
Sarazin, C. L., Irwin, J. A., \& Bregman, J. N. 2001, \apj, 556, 553

\bibitem[]{}
Satyapal, S., Vega, D., Dudik, R. P., Abel, N. P., \& Heckman, T. 2008, \apj, 677, 926

\bibitem[]{}
Schawinski, K., \etal 2010, \aj, in press (arXiv:1001.3141)

\bibitem[]{}
Schawinski, K., Virani, S., Simmons, B., Urry, C. M., Treister, E., Kaviraj, S., \& Kushkuley, B. 2009, \apjl, 692, L19

\bibitem[]{}
Schawinski, K., \etal 2007, \mnras, 382, 1415

\bibitem[]{}
Seth, A., \etal 2010, AIP Conf. Series, in press (arXiv:1002.0824)

\bibitem[]{}
Seth, A.,  Ag\"ueros, M., Lee, D., \& Basu-Zych, A. 2008, \apj, 678, 116

\bibitem[]{}
Shankar, F., Weinberg, D. H., \& Miralda-Escud\'e, J. 2009, \apj, 690, 20

\bibitem[]{}
Silverman, J. D., et al. 2008, \apj, 675, 1025

\bibitem[{{Sivakoff} {et~al.}(2007){Sivakoff}, {Jord{\'a}n}, {Sarazin},
  {Blakeslee}, {C{\^o}t{\'e}}, {Ferrarese}, {Juett}, {Mei}, \&
  {Peng}}]{Siv++07}
{Sivakoff}, G.~R., {Jord{\'a}n}, A., {Sarazin}, C.~L., {Blakeslee}, J.~P.,
  {C{\^o}t{\'e}}, P., {Ferrarese}, L., {Juett}, A.~M., {Mei}, S., \& {Peng},
  E.~W. 2007, \apj, 660, 1246

\bibitem[]{}
Sivia, D. S. 2006, {\it Data Analysis, A Bayesian Tutorial}, Oxford University Press (Oxford)

\bibitem[{{Soltan}(1982)}]{Sol82}
{Soltan}, A. 1982, \mnras, 200, 115

\bibitem[{{Soria} {et~al.}(2006a){Soria}, {Graham}, {Fabbiano}, {Baldi},
  {Elvis}, {Jerjen}, {Pellegrini}, \& {Siemiginowska}}]{Sor++06}
{Soria}, R., {Graham}, A.~W., {Fabbiano}, G., {Baldi}, A., {Elvis}, M.,
  {Jerjen}, H., {Pellegrini}, S., \& {Siemiginowska}, A. 2006a, \apj, 640, 143

\bibitem[{{Soria} {et~al.}(2006b){Soria}, {Graham}, {Fabbiano}, {Baldi},
  {Elvis}, {Jerjen}, {Pellegrini}, \& {Siemiginowska}}]{Sor++06}
{Soria}, R., {Graham}, A.~W., {Fabbiano}, G., {Baldi}, A., {Elvis}, M.,
  {Jerjen}, H., {Pellegrini}, S., \& {Siemiginowska}, A. 2006b, \apj, 640, 126

\bibitem[]{}
Springel, V., Di Matteo, T., \& Hernquist, L. 2005, \mnras, 361, 776

\bibitem[]{}
Terashima, Y., \& Wilson, A. S. 2003, \apj, 583, 145

\bibitem[{{Tremaine} {et~al.}(2002){Tremaine}, {Gebhardt}, {Bender}, {Bower},
  {Dressler}, {Faber}, {Filippenko}, {Green}, {Grillmair}, {Ho}, {Kormendy},
  {Lauer}, {Magorrian}, {Pinkney}, \& {Richstone}}]{Tre++02}
{Tremaine}, S., {Gebhardt}, K., {Bender}, R., {Bower}, G., {Dressler}, A.,
  {Faber}, S.~M., {Filippenko}, A.~V., {Green}, R., {Grillmair}, C., {Ho},
  L.~C., {Kormendy}, J., {Lauer}, T.~R., {Magorrian}, J., {Pinkney}, J., \&
  {Richstone}, D. 2002, \apj, 574, 740

\bibitem[{{Valluri} {et~al.}(2005){Valluri}, {Ferrarese}, {Merritt}, \&
  {Joseph}}]{Val++05}
{Valluri}, M., {Ferrarese}, L., {Merritt}, D., \& {Joseph}, C.~L. 2005, \apj,
  628, 137

\bibitem[]{}
Van Wassenhoven, S., Volonteri, M., Walker, M. G., \& Gair, J. R. 2010, MNRAS submitted (arXiv:1001.5451)

\bibitem[{{Volonteri} {et~al.}(2008{\natexlab{a}}){Volonteri}, {Haardt}, \&
  {Gultekin}}]{VHG07}
{Volonteri}, M., {Haardt}, F., \& {G\"ultekin}, K.  2008, \mnras\, 384, 1387

\bibitem[]{}
Wehner, E. H., \& Harris, W. E., 2006, \apj, 644, L17

\bibitem[{{Woo} \& {Urry}(2002)}]{W+U02}
{Woo}, J.-H., \& {Urry}, C.~M. 2002, \apj, 579, 530

\bibitem[]{}
Yu, Q., \& Tremaine, S. 2002, \mnras, 335, 965

\bibitem[{{Zhao} {et~al.}(2005){Zhao}, {Grindlay}, {Hong}, {Laycock}, {Koenig},
  {Schlegel}, \& {van den Berg}}]{Zha++05}
{Zhao}, P., {Grindlay}, J.~E., {Hong}, J.~S., {Laycock}, S., {Koenig}, X.~P.,
  {Schlegel}, E.~M., \& {van den Berg}, M. 2005, \apjs, 161, 429

\bibitem[]{}
Zhang, W. M., Soria, R., Zhang, S. N., Swartz, D. A., \& Liu, J. F. 2009, \apj,  699, 281


\end{thebibliography}
\end{document}